\documentclass[12pt]{article}
\usepackage{graphicx,amssymb,amsmath,amsthm,cite}
\usepackage{algorithmic,setspace}
\usepackage[ruled]{algorithm}
\usepackage{color,multirow}
\usepackage{verbatim}
\usepackage{float}
\def\vRt{\til{\mathbf{R}}}
\newcommand{\la}{\langle}
\newcommand{\ra}{\rangle}
\newcommand{\eps}{\epsilon}
\newcommand{\Spann}{{\mbox{\rm{span}}}}
\newcommand{\til}{\tilde}

\newcommand{\vc}{\mathbf{c}}

\def\be{\begin{equation}}
\def\ee{\end{equation}}
\def\ben{\begin{eqnarray}}
\def\een{\end{eqnarray}}
\def\vI{\mathbf{I}}
\def\vL{\mathbf{L}}

\def\vG{\mathbf{G}}
\def\vB{\mathbf{B}}
\def\vW{\mathbf{W}}
\def\vU{\mathbf{U}}
\def\vR{\mathbf{R}}
\def\vD{\mathbf{D}}
\def\vA{\mathbf{A}}
\def\vd{\mathbf{d}}
\def\vg{\mathbf{g}}

\def\vp{\mathbf{p}}
\def\vh{\mathbf{h}}
\def\F{\mathrm{3D}}

\def\Dc{\D_{\rm{C}}}
\def\Ds{\D_{\rm{S}}}

\def\Dwd{\D_{\rm{wd}}}
\def\tDwd{\tD_{\rm{wd}}}
\def\Dpd{\D_{\rm{pd}}}
\def\tDpd{\tD_{\rm{pd}}}
\def\DLw{\D_{\rm{Lw}}}
\def\DLp{\D_{\rm{Lp}}}

\def\SNR{\mathrm{SNR}}
\def\SRDD{\mathrm{SR}_{\mathrm{2D}}}
\def\SRD{\mathrm{SR}_{\mathrm{3D}}}
\def\SR{\mathrm{SR}}
\def\SRm{\overline{\mathrm{SR}}}
\def\PSNR{\mathrm{PSNR}}
\def\IM{\mathrm{Imax}}

\def\MSE{\mathrm{MSE}}
\def\op{\hat{\mathrm{P}}}

\def\PSNR{\mathrm{PSNR}}
\def\MSE{\mathrm{MSE}}
\def\Nq{N_b}
\def\Nb{N_b}
\def\Nx{N_x}
\def\Nx{N_x}
\def\Nxy{\Nx \times \Ny}
\def\Mx{M_x}
\def\My{M_y}

\def\vdx{{\vd}^x}

\def\vdy{{\vd}^y}
\def\vdz{{\vd}^z}

\def\llx{{\ell^{x}}}
\def\lly{{\ell^{y}}}
\def\llz{{\ell^{z}}}

\def\D{\mathcal{D}}
\def\tD{\til{\mathcal{D}}}

\def\R{\mathbb{R}}

\def\V{\mathbb{V}}

\def\kpl{k_{p,l}}
\def\kq{k_q}
\def\Nx{N_x}
\def\Ny{N_y}
\def\Nz{N_z}
\def\Nxy{\Nx \times \Ny}
\def\Nxyz{\Nx \times \Ny \times \Nz}

\def\Mx{M_x}
\def\My{M_y}
\def\Mz{M_z}
\title{Sparse Representation of 3D Images for 
 Piecewise Dimensionality Reduction 
 with High Quality Reconstruction}
\author{Laura Rebollo-Neira and 
Daniel Whitehouse\\%\footnotemark[2]}
Mathematics Department\\
Aston University\\
B3 7ET, Birmingham, UK}
\begin{document}
\maketitle
\baselineskip = 1.5\baselineskip
\begin{abstract}
Sparse representation of 3D images is considered 
within the context of data reduction. The goal is to produce high quality approximations of 3D images using fewer elementary components than the number of intensity points in the 3D array. This is achieved by means of a highly redundant dictionary and a dedicated pursuit strategy especially designed for low memory requirements.  The benefit of the proposed framework is illustrated in the first instance by demonstrating the gain in dimensionality reduction obtained when approximating true color images as very thin 3D arrays, instead of performing an independent  channel by channel approximation. The full power of the approach is further exemplified by producing high quality approximations of hyper-spectral images with a reduction of up to 371 times the number of data points in the representation. 

 \end{abstract}

{\bf{Keywords:}} Image representation with Dictionaries;
Greedy Pursuit Algorithms.

\section{Introduction}
Sparse representation of 2D images 
has been a subject of extensive research in the 
last fifteen years \cite{WMM10, Ela10, ZXY15}. 
Applications which benefit from sparsity 
range from image restoration \cite{MES08,ZSL13} 
and classification \cite{GHO18,GYZ18,GWY18}   
to feature extraction \cite{WYG09,YLY12}  
 and super resolution reconstructions \cite{YWH10,ZLY15}. 
 While sparse representation of 
 3D arrays has received 
 less attention, the advantage of modeling these arrays
 as a superposition of 3D elementary 
  components is recognized in previous publications 
 \cite{DFK07,CC13,CML15,DYK17}.

At present,  the most widely  
used  multichannel images in every day life  are
true color images.
 The simplest way of sparsely representing these images is 
channel by channel, or adding constraints of 
 correlation across colors \cite{MES08,MM17}.
However, as demonstrated in this work, 
 sparsity in the representation of
true color images can increase substantially 
if the approximation is realized by means of
3D elements taken from a highly redundant dictionary. 
The effect is of course more pronounced  
 for arrays involving more channels, such as
 hyper-spectral images. 

From a practical view point, 
the current drawbacks of 3D sparse modeling using 
a large dictionary are (i) storage 
requirements and (ii) the complexity of the concomitant 
calculations. In this paper we propose a method 
 which, by addressing (i) leaves room for 
possible high performance implementations using 
Graphics Processing Unit (GPU) programming. 
While the approach is illustrated using 
 Central Processing Unit (CPU) programming, 
 the storage requirements 
are shown to fit within 48Kb's of fast access shared memory 
of a GPU when the approximation of a 3D image is realized 
with a partition  block size of $8 \times 8 \times 8$ and 
with a separable dictionary of redundancy 125.

The main contributions of the paper are listed below.
\begin{itemize}
\item
The low memory implementation of the 
Orthogonal Matching Pursuit (OMP) 
strategy, called  Self Projected Matching Pursuit (SPMP) 
\cite{RNB13}  
is dedicated to operating in 3D (SPMP3D) with  
separable dictionaries.
This technique delivers an iterative solution to the 
 3D least squares problem which requires  much less storage 
 than direct linear algebra methods. 
It could therefore be also applied with any other   
 of the pursuit strategies that include a least 
 squares step \cite{DTD06,NT09, EKB10, CC13, RNMB13}. 
\item
The C++ MEX file for the SPMP3D method has  
been made available on a dedicated website \cite{webpage}.
All the scripts for reproducing the results of the 
paper in the MATLAB environment 
have also been placed on that website.
\item
Remarkable reduction in the dimensionality of 
 the representation of true color images and 
hyper-spectral images, with 
 high quality reconstruction, is demonstrated using 
 highly redundant and highly coherent
separable dictionaries.
\end{itemize}
 The results suggest that the  method may be of 
assistance to 
 image processing applications which rely
on a transformation for data reduction as a first step
of further processing. For examples of relevant 
applications we refer to 
 \cite{TWH15,NLS16,GLZ17,ZCL18,LPN18}. 
\section{Notational Convention}
\label{notation}
$\R$ represents the set of real numbers.
Boldface letters are used to indicate Euclidean vectors, 
2D and 3D arrays. 
Standard mathematical fonts indicate components,
e.g., $\vd \in \R^N$ is a vector of components
$d(i)\in \R,\, i=1,\ldots,N$. The  elements of a 
 3D array $\vI \in \R^{\Nxyz}$ are indicated as
$I(i,j,m),\,i=1,\ldots, \Nx, \, j=1,\ldots,\Ny , \, m=1,\ldots,\Nz$.  
Moreover, for each $m$-value $\vI_m \in  \R^{\Nxy}$ stands for the 
2D array of elements $I_m(i,j)=I(i,j,m),\,i=1,\ldots,\Nx,\, j=1,\ldots,\Ny$, which, when not leaving room for 
ambiguity will also be represented as 
$I(:,:,m)$. The transpose of a matrix, $\vG$ say, is 
indicated as $\vG^\top$.

The inner product between 3D arrays, say $\vI \in  \R^{\Nxyz}$ and 
$\vG \in \R^{\Nxyz}$, is given as: 
$$\la \vG, \vI \ra_{\F}= \sum_{i=1}^{\Nx} \sum_{j=1}^{\Ny} \sum_{m=1}^{\Nz} G(i,j,m)  I(i,j,m).$$
For $\vG \in \R^{\Nxyz}$ with tensor product structure, i.e. 
for $\vG = \vg^x \otimes \vg^y \otimes \vg^z$, 
with $\vg^x \in \R^{\Nx}, \vg^y \in \R^{\Ny}$ and $\vg^z \in \R^{\Nz}$, we further have
\be
\label{inp3d}
\la \vG, \vI \ra_{\F}= \sum_{m=1}^{\Nz} \la \vg^x, \vI_m \vg^y \ra g^z(m)
= \la \vp, \vg^z \ra,
\ee
where for each value of $m$ the vector $\vI_m \vg^y$ in  $\R^{\Nx}$ arises by the standard matrix-vector multiplication rule and  $\vp \in \R^{\Nz}$ is given by its components 
$p(m)=\la \vg^x, \vI_m \vg^y \ra,\, m=1,\ldots,\Nz$. Note that 
$\la \vp, \vg^z \ra$ indicates the Euclidean inner product in 1D, i.e. 
$$\la \vp, \vg^z \ra= \sum_{m=1}^{\Nz} p(m) g^z(m).$$
The definition $\eqref{inp3d}$ induces the norm
$\|\vI\|_{\F}= \sqrt{\la \vI, \vI \ra_{\F}}$.
\section{Sparse Representation of Multi-channel Images}
Suppose that a 3D image, given as an array 
$\vI \in \R^{\Nxyz}$ of intensity 
pixels, is to be approximated by the linear
decomposition
\be
\vI^k= \sum_{n=1}^k c(n) \vD_{\ell_n},
\label{atom}
\ee
where each $c(n)$ is a scalar and each $\vD_{\ell_n}$ 
is an element of $\R^{\Nxyz}$ to be selected from a 
set, $\D=\{\vD_n\}_{n=1}^M$, 
called a `dictionary'.

A sparse approximation of $\vI\in \R^{\Nxyz}$  is 
an approximation of the form \eqref{atom} such that the 
 number $k$
of elements in the decomposition is significantly smaller
than ${N=\Nx \Ny \Nz}$.  
The terms in the decomposition \eqref{atom}
are taken from a large 
redundant dictionary, from where the 
elements $\vD_{\ell_n}$ in \eqref{atom}, called `atoms', 
are chosen according to an optimality criterion.

Within the redundant  dictionary framework for 
approximation, the problem of finding the sparsest decomposition of a given multi-channel image
can be formulated as follows: 
{\em{Given an image and a dictionary, 
approximate the image  by the `atomic decomposition' 
\eqref{atom}
such that the number $k$ of atoms is minimum.}} 
Unfortunately, the numerical
minimization of the number of terms to produce an 
approximation up to a desired error, 
involves a combinatorial problem for
exhaustive search. Hence, the solution 
 is intractable.
Consequently, instead of looking for the
sparsest solution, one looks for a
`satisfactory solution',  i.e.,
a solution such that the number of $k$-terms 
in \eqref{atom}
is considerably smaller than the image dimension.  
For 2D images this can be effectively achieved by
 greedy pursuit strategies in the line of the 
Matching Pursuit (MP) \cite{MZ93} and  
OMP \cite{PRK93}
methods, if dedicated to 2D separable dictionaries 
\cite{RNBCP12,RNB13,CC13,LRN17}. 
Within a tensor product framework the consideration 
of OMP in 3D is natural.

Let's assume that a 3D dictionary is obtained as the 
tensor product 
 $\D=\D^x \otimes \D^y \otimes \D^z$ of three
1D dictionaries
$\D^x =\{\vd^x_m \in \R^{\Nx}\}_{m=1}^{\Mx}$,  
$\D^y =\{\vd^y_m \in \R^{\Ny}\}_{m=1}^{\My}$, and   
$\D^z =\{\vd^z_m \in \R^{\Nz}\}_{m=1}^{\Mz}$, with 
$\Mx \My \Mz= M$. For computational purposes the 1D 
dictionaries are stored as three matrices
$\vD^x \in \R^{\Nx \times \Mx}$, 
$\vD^y \in \R^{\Ny \times \My}$ and
$\vD^z \in \R^{\Nz \times \Mz}$.
Suppose now that a 3D array $\vI \in 
\R^{\Nx \times \Ny \times \Nz}$ is to  be 
 approximated by an {{atomic decomposition}} of the 
form
\be
\label{atoq}
\vI^{k}= \sum_{n=1}^{k} 
c(n) \vd^x_{\ell^{x}_n} \otimes \vd^y_{\ell^{y}_n} \otimes \vd^z_{\ell^{z}_n},
\ee
where for $n=1,\ldots,k$ the atoms $\vd^x_{\ell^{x}_n}$, 
$\vd^y_{\ell^{y}_n}$ and $\vd^z_{\ell^{z}_n}$ are 
selected from the given 1D dictionaries. 
The common step of the techniques we
consider for constructing
approximations of the form \eqref{atoq} is the
stepwise selection of the atoms in the 
atomic decomposition. 
On setting $k=1$ and $\vI^0=0$ at
iteration $k$ the algorithm selects the indices
$\ell^{x}_{k}$, $\ell^{y}_{k}$ and 
$\ell^{z}_{k}$ as follows
\be
\label{select}
\ell^{x}_{k},\ell^{y}_{k}, 
\ell^{z}_{k}= \operatorname*{arg\,max}_{\substack{n=1,\ldots,\Mx\\
i=1,\ldots,\My\\
s=1,\ldots,\Mz}} \left |\la \vd^{x}_n \otimes \vd^{y}_i \otimes \vd^{z}_s ,\vR^{k-1} \ra_{\F} \right|, 
\ee
with $\vR^{k-1}= \vI - \vI^{k-1}$. 
It is the determination 
of the coefficients $c(n),\,n=1,\ldots,k$ in 
\eqref{atoq} that   
gives rise to pursuit strategies which go with 
different names. 
\subsection{Matching Pursuit in 3D (MP3D)}
The MP approach in 3D would simply calculate 
the coefficients in 
\eqref{atoq} as 
\be
\label{cmp}
c(n)= \la \vd^x_{\ell^{x}_n} \otimes  \vd^y_{\ell^{y}_n} \otimes \vd^z_{\ell^{z}_n} ,\vR^{n-1} \ra_{\F}, \quad 
n=1,\ldots,k.
\ee
The main drawback of the MP method  is that it may select linearly 
dependent atoms. Moreover, that approximation  is not 
stepwise optimal because at iteration $k$ the 
coefficients \eqref{cmp} 
do not minimize the norm of the residual error. 
 The pursuit strategy that overcomes 
these limitations is the so called 
OMP \cite{PRK93}.
\subsection{Orthogonal Matching Pursuit in 3D}
The implementation of OMP in 3D (OMP3D) we 
describe here is the 3D 
extension of the implementation of OMP in 2D 
 given in \cite{RNBCP12}. An alternative 
algorithm called Kronecker-OMP, which is based on 
the Tucker representation of a tensor, is discussed in 
\cite{CC13}. 
Our algorithm  is based on adaptive 
biorthogonalization and 
Gram-Schmidt orthogonalization procedures, as 
proposed in \cite{RNL02} for the one dimensional case. 

In order to ensure the coefficients 
$c(n),\,n=1,\ldots,k$ involved in
\eqref{atoq}
are such that
$\|\vR^{k}\|_{\F}^2= \la\vR^{k}, \vR^{k} \ra_{\F} $ is minimum, the decomposition \eqref{atoq} should  
 fulfill that
\be
\label{atop}
\vI^{k}= \sum_{n=1}^{k}
c(n) \vd^x_{\ell^{x}_n} \otimes \vd^y_{\ell^{y}_n} \otimes \vd^z_{\ell^{z}_n}= \op_{\V_k} \vI,
\ee
where $\op_{\V_k}$ is the orthogonal
projection operator
onto $\V_k=\Spann\{\vdx_{\ell^x_n} \otimes \vdy_{\ell^y_n} \otimes \vdz_{\ell^z_n} \}_{n=1}^k$.
This is ensured by requiring that 
$\vR^{k}= \vI- \op_{\V_k} \vI$,
where $\op_{\V_k}$ is the orthogonal projection operator
onto $\V_k=\Spann\{\vdx_{\ell^x_n} \otimes \vdy_{\ell^y_n} \otimes \vdz_{\ell^z_n} \}_{n=1}^k$.
The required representation of $\op_{\V_k}$ is of 
 the form
$\op_{\V_k} \vI  = \sum_{n=1}^k \vA_n \la \vB_n^k, \vI \ra_{\F} $,
where each $\vA_n \in \R^{\Nxyz}$ is an array with the
selected atoms $\vA_n= \vdx_{\ell^x_n} \otimes \vdy_{\ell^y_n} \otimes \vdx_{\ell^z_n}$.  The
 concomitant biorthogonal reciprocal 
set $\vB_n^k,\,n=1,\ldots,k$ comprises the unique elements
 of $\R^{\Nxyz}$ satisfying the conditions:
 \begin{itemize}
 \item
 [i)]$\la \vA_n, \vB_m^k \ra_{\F}=\delta_{n,m}= \begin{cases}
 1 & \mbox{if}\, n=m\\
 0 & \mbox{if}\, n\neq m.
 \end{cases}$
\item
[ii)]${\V_k}= \Spann\{\vB_n^k\}_{n=1}^k.$
\end{itemize}
Thus, the coefficients $c(n),\,n=1,\ldots,N$ in \eqref{atop} 
which guarantee minimum norm of the residual error 
are calculated as
$$c(n)=\la \vB_n^k, \vI \ra_{\F},\quad n=1,\ldots,k.$$
The required arrays $\vB_n^k,\,n=1,\ldots,k$ should be 
 upgraded and updated to account for each newly 
selected atom. 
Starting from 
$k=1,\vR^{0}=\vI$, $\vB_1^1=\vW_1=
\vA_1= \vdx_{\ell^x_1} \otimes \vdy_{\ell^y_1} \otimes \vdz_{\ell^z_1}$, 
where
$$\ell^{x}_{1},\ell^{y}_{1},
\ell^{z}_{1}= \operatorname*{arg\,max}_{\substack{n=1,\ldots,\Mx\\
i=1,\ldots,\My\\
s=1,\ldots,\Mz}} \left |\la \vd^{x}_n \otimes \vd^{y}_i \otimes \vd^{z}_s ,\vR^{k-1} \ra_{\F} \right|,
$$
at iteration $k+1$ the indices 
$\ell^{x}_{k+1},\ell^{y}_{k+1},
\ell^{z}_{k+1}$ corresponding to the new 
 atom $\vA_{k+1}= \vdx_{\ell^x_{k+1}} \otimes \vdy_{\ell^y_{k+1}} \otimes \vdz_{\ell^z_{k+1}}$ are selected as in 
\eqref{select}. The required 
reciprocal set $\vB_n^{k+1},\,n=1,\ldots,k+1$ is 
adapted and upgraded by extending the recursion formula given in \cite{RNL02} as follows.
\be
 \begin{split}
 \vB_n^{k+1}&= \vB_n^k - \vB_{k+1}^{k+1}\la \vA_{k+1}, \vB_n^k \ra_{\F},\quad n=1,\ldots,k,\\
 \text{where}\\
 \vB_{k+1}^{k+1}&= \vW_{k+1}/\|\vW_{k+1}\|_{\F}^2,\\
  \text{with} \\
 \vW_{k+1}&= \vA_{k+1} - \sum_{n=1}^k \frac{\vW_n}
 {\|\vW_n\|_{\F}^2} \la \vW_n, \vA_{k+1}\ra_{\F}, \nonumber
 \end{split}
\ee
including, for numerical accuracy, the re-orthogonalization 
step: 
\be
\label{RGS}
\vW_{k+1} \leftarrow   \vW_{k+1} - \sum_{n=1}^k \frac{\vW_n}
 {\|\vW_n\|_{\F}^2} \la \vW_n, \vW_{k+1}\ra_{\F}. \nonumber
\ee
%With the arrays $\vB_n^k,\,n=1,\ldots,k$ constructed 
%as above the required coefficients
%in \eqref{atop} are obtained from the
%inner products
%$$c_n= \la \vB_n^k, \vI \ra_{\F},\, n=1,\ldots,k.$$ 
 Although the image approximation is carried out 
by partitioning the images into  
relatively small 3D blocks,  memory 
requirements of the OMP3D method are high. 
Indeed, the above are $2(k+1)$ nonseparable arrays each of 
dimension $N=\Nx \Ny \Nz$ which need to be stored in 
double  precision.  
Hence, we consider next a low memory implementation of the 
orthogonal projection step, which avoids having to 
store the arrays $\vW_n,\,n=1,\ldots,k$ and $\vB_n^k,\,n=1,\ldots,k$ and fully exploits the separability of the 
dictionary.
\subsection{Self Projected  Matching Pursuit in 3D (SPMP3D)}
The Self Projected Matching Pursuit (SPMP) methodology 
was introduced  in \cite{RNB13} and conceived to be 
used with separable dictionaries in 2D (SPMP2D). 
Because the technique is 
based on calculations of inner products, 
it can be easily extended to operate in 3D (SPMP3D). 

Suppose that at iteration $k$ the selection process has 
chosen the atoms labeled by the triple of indices 
$\{\ell^{x}_n ,\ell^{y}_n,\ell^{z}_n\}_{n=1}^{k}$ 
and let $\til{\vI}^{k}$ be the atomic decomposition
\be
\label{atoq2}
\til{\vI}^{k}= \sum_{n=1}^{k}
a(n)\vd^x_{\ell^{x}_n} \otimes \vd^y_{\ell^{y}_n}
B\otimes \vd^z_{\ell^{z}_n},
\ee
where the coefficients $a(n), \, n=1,\ldots,k$
are arbitrary numbers.
Every array ${\vI} \in 
\R^{\Nxyz}$ can be expressed as
\be
{\vI}= \til{\vI}^{k} + \til{\vR}.
\ee
For $\til{\vI}^{k}$ to be the optimal representation
of ${\vI}$ in
$\V_{k}= \Spann\{\vd^x_{\ell^{x}_n} \otimes 
\vd^y_{\ell^{y}_n} \otimes \vd^z_{\ell^{z}_n}\}_{n=1}^{k}$, in the sense of minimizing the
norm of the residual $\vRt$, it should be true that
$\op_{\V_{k}} \vRt=0$. The SPMP3D method fulfills this
property by approximating $\vRt$ in $\V_{k}$, via the
MP method, and subtracting that component from $\vRt$.
The following algorithm describes the  whole
procedure.
%Set $\Ga^x_0=\{\emptyset\}$, $\Ga^y_0=\{\emptyset\}$, 
%$\Ga^z_0=\{\emptyset\}$ $\vI^0=0$ and $\vR^0=\vI$.
Starting from $k=0$ and $\vR^0=\vI$, 
at each iteration, implement the steps below.
\begin{itemize}
\item[i)]
 Increase $ k \leftarrow k+1$ and apply the criterion \eqref{select} for selecting
 the  triple of indices 
$(\ell^{x}_{k},\ell^{y}_{k},\ell^{z}_{k})$.  
 Save this triple in the array 
 $L(k,1:3)=(\ell^{x}_{k},\ell^{y}_{k},\ell^{z}_{k}),$
 and set $$c(k)=\la \vd^x_{\ell^{x}_k} \otimes 
\vd^y_{\ell^{y}_k} \otimes \vd^z_{\ell^{z}_k}, \vR^{k-1}\ra_{\F}$$ 
and implement the update of the residue 
$\vR^k=\vR^{k-1} - c(k) \vd^x_{\ell^{x}_{k}} \otimes 
\vd^y_{\ell^{y}_{k}}  \otimes  \vd^z_{\ell^{z}_{k}}$ 
as follows:\\
For $s=1,\ldots\Nz$ calculate 
 $$\Delta \vR^k(:,:,s) = \vR^{k-1}(:,:,s) - 
 c(k) \vd^x_{\ell^{x}_{k}} 
(\vd^y_{\ell^{y}_{k}})^\top  d^z_{\ell^{z}_{k}}(s),$$
 to   
update  $\vR^k$ as 
$$\vR^k= \vR^{k-1} - \Delta \vR^k.$$ 
\item[ii)]
Given the indices 
$L(n,1:3)=(\ell^{x}_{n},\ell^{y}_{n},\ell^{z}_{n}),\, n=1,\ldots,k$ of the previously selected atoms, and a 
tolerance  $\eps$ for the projection
error, realize the orthogonal projection 
up to that error as follows.
Set $j=1$, $\vRt^{0}=\vR^{k}$  and 
at iteration $j$ apply the steps a) - c)  below.
\begin{itemize}
\item [a)]
 For $n=1,\ldots,k$ evaluate 
\be
\label{alpha}
\alpha(n)=
\la \vd^{x}_{\ell^{x}_{n}} 
\otimes \vd^{y}_{\ell^{y}_{n}} \otimes \vd^{z}_{\ell^{z}_{n}}, \vRt^{j-1}\ra_{\F}, 
\ee
and single out the value $k^\ast$ such that
\be
\label{ast}
k^\ast =\!\operatorname*{arg\,max}_{\substack{n=1,\ldots,k}} \! | \alpha(n)|.
\ee
The value $k^\ast$  signalizes  the indices   
$\ell^{x}_{k^\ast},\ell^{y}_{k^\ast},\ell^{z}_{k^\ast}$
corresponding to the already selected atoms with maximum 
  correlation with the residual $\vRt^{j-1}$.
\item [b)] If 
  $|\alpha(k^\ast)| < \eps $ 
stop. Otherwise 
 update the coefficient $$c(k^\ast) \leftarrow 
 c(k^\ast) +  \alpha(k^\ast) $$ and 
 for $s=1,\ldots,\Nz$  evaluate
$$\Delta \vRt^j(:,:,s)=  
 \alpha(k^\ast)\, \vd^x_{\ell^{x}_{k^\ast}} (\vd^y_{\ell^{y}_{k^\ast}})^\top d^z_{\ell^{z}_{k^\ast}}(s)$$
to update the residual $\vRt^j$  as
$$\vRt^j= \vRt^{j-1} - \Delta \vRt^j.$$
This step subtracts from the residual a
component in $\V_{k}$ and add that component to the 
approximation $\til{\vI}^{k}$ 
\item [c)] Increase  $j \leftarrow j+1$
 and repeat the steps a) - c) to keep subtracting 
components of $\vRt^j$ in $\V_{k}$ until   
 iteration, $J$ say, for which the stopping criterion b) 
is met. This criterion indicates that,  
up to  tolerance $\epsilon$,  the 
 residual has no component in  $\V_{k}$ so that one can set 
  $\vR^k= \til{\vR}^{J-1}$.
\end{itemize}
Continue with 
steps i) and ii) to keep enlarging 
 $\V_{k}$  until, for a required 
tolerance error $\rho$, the condition 
$\| \vR^k\|_{3D} < \rho$ is reached. 
\end{itemize}
{\em{Remark 1:}}
 For each fixed value $k$ the  rate of convergence 
$$\lim_{j \to \infty} \vI - \til{\vR}^j= \op_{\V_{k}} \vI$$
 through the steps 
a) - c) above is given 
 in \cite{RNRS17} for the one dimensional case. The proof
for 3D is identical to that proof, because a 
 3D array can be represented as a long 1D vector. 
What varies is the implementation. A vectorized 
version of the algorithm would not be applicable 
in this context.
\subsubsection*{Implementation details}
The bulk of the computational burden in the SPMP3D method  
 lies in the realization of the selection 
of atoms \eqref{select}. Algorithm~\ref{IP3D} outlines a  
 procedure implementing the process. It should be
 stressed once 
again  that 
the algorithm is designed to use as little memory 
as possible, rather than to reduce complexity. 
\newcounter{myalg}
\begin{algorithm}[!htp]
\refstepcounter{myalg}
\begin{algorithmic}
\caption{Implementation of the selection of atoms 
(c.f. \eqref{select})\newline
Procedure $[ {\alpha},\llx, \lly, \llz]=\text{Sel3DAtom}(\vR,\vD_x,\vD_y,\vD_z)$}
\label{IP3D}
\STATE{{\bf{Input:}}\, 3D array $\vR$, 
 matrices  $\vD_x$,  $\vD_y$ $\vD_z$ the columns 
of which are the 
atoms in the corresponding dictionaries.}
\STATE{{\bf{Output:}}\, selected indices $\llx, \lly, \llz$,  and ${\alpha}= \la \vd^x_{\ell^{x}} \otimes \vd^y_{\ell^{y}} \otimes  \vd^z_{\ell^{z}}, \vR \ra_{\F}$}
%\STATE{$[\Nx M_x]=\text{size}(\vD_x)$}
%\STATE{$[\Ny M_y]=\text{size}(\vD_y)$}
\STATE{\COMMENT{Initiate the algorithm}}
\STATE{$(\Nz,\Mz)=\text{size}(\vD_z)\;,\Mx=\text{size}(\vD_x,2);\,\My=\text{size}(\vD_y,2)\,$}
\STATE{$q=\text{zeros}(\Mx,\My)$}
\STATE{${\alpha}=0$}
\FOR {$m=1:\Mz$}
\STATE{$q(:,:)=0$}
\FOR {$s=1:\Nz$}
\STATE{$q(:,:)=q(:,:)+\vD_x^\top R(:,:,s)\vD_y d^z_m(s)$}
 \ENDFOR

\COMMENT{Realize \eqref{select} by finding the partial 
 maximum, and its argument, for each $m$-plane} 

\STATE{$[l_1,l_2,\til{q}]=\text{max}(|q(:,:)|)$}
\IF {$\til{q} > {\alpha}$}
\STATE{${\alpha} = \til{q};\, \llx=l_1;\lly=l_2;\, \llz=m$}
\ENDIF
\ENDFOR
\end{algorithmic}
\end{algorithm}
At iteration $k$ the outputs of Algorithm~\ref{IP3D} are 
saved as $c(k)={\alpha}$ and $L(k,1:3)=(\llx, \lly, \llz)$.
The implementation details for selecting the 
 triple of indices at the projection step are given in 
Algorithm~\ref{SelTrip}. This is used in 
Algorithm~\ref{Proj} for the realization of the actual 
projection to recalculate the coefficients in the atomic 
decomposition.
\begin{algorithm}[!htp]
\refstepcounter{myalg}
\begin{algorithmic}
\caption{Selection of the triple of indices from the reduced dictionary
(c.f. \eqref{ast})\newline
Procedure 
$[\alpha^{\ast},k^{\ast}]$=\text{SelTrip}$(\vR, \vD_x,\vD_y,\vD_z,\vL)$}
\label{SelTrip}
\STATE{{\bf{Input:}}\, As in Algorithm~\ref{IP3D} plus 
the array $\vL$,  
 with the  triple of indices 
$L(n,1:3)=(\ell^{x}_{n},\ell^{y}_{n},\ell^{z}_{n}),\, n=1\ldots k$}
\STATE{{\bf{Output:}}\,$k^{\ast}$
and the corresponding values of $\alpha$ 
(c.f. \eqref{ast}) to update 
the coefficients and residual}
\STATE{\COMMENT{Initiate the algorithm}}
\STATE{$\alpha^{\ast}=0$}
\FOR {$n=1:k$}
\STATE{$p=0$}
\FOR {$s=1:\Nz$}
\STATE{$p=p+(\vd^x_{\ell^{x}_{n}})^\top R(:,:,s) \vd^y_{\ell^{y}_{n}} d^z_{\ell^{z}_{n}}(s)$}
 \ENDFOR
\IF{$|p| > |\alpha^{\ast}|$}
\STATE{$k^{\ast}=n$ 
 and $\alpha^{\ast}=p$}
\ENDIF
\ENDFOR
\end{algorithmic}
\end{algorithm}

\begin{algorithm}[!ht]
\refstepcounter{myalg}
\begin{algorithmic}
\caption{Implementation of the self projection steps 
a) - c).\newline
Procedure 
$[\til{\vR}, \bar\vc]$=\text{Proj3D}($\vR,\vD_x,\vD_y,\vD_z, \vL, \vc, \eps, \text{MaxJ}$).}
\label{Proj}
\STATE{{\bf{Input:}}\, As in Algorithm \ref{SelTrip}, plus 
the coefficients of the atomic decomposition $\vc$, 
a tolerance parameter $\eps$ 
for the numerical error of the projection, 
and a maximum number of permitted iterations, MaxJ.}
\STATE{{\bf{Output:}}\, Orthogonal residual $\til{\vR}$. 
Coefficients $\til{\vc}$ of the optimized 
atomic decomposition.}
%\STATE{\COMMENT{Initiate the algorithm}}
\FOR {$j=1:\text{MaxJ}$}
\STATE{\COMMENT{Selection of atoms using Algorithm~\ref{SelTrip}}}
\STATE{$[\alpha^\ast, k^\ast]$=\text{SelTrip}($\vR,\vD_x,\vD_y,\vD_z, \vL$)}
\STATE{\COMMENT{Check stopping criterion}}
\IF{$|\alpha^\ast| < \eps$} 
\STATE{stop} 
\ENDIF
\STATE{\COMMENT{Update the coefficients}}
\STATE{$c(k^\ast)  \leftarrow  c(k^\ast)+ \alpha^\ast$}
\STATE{\COMMENT{Update the residual}}
\FOR {$s=1: \Nz$}
\STATE{$\vR(:,:,s) \leftarrow \vR(:,:,s)- \alpha^\ast(\vd^x_{\ell^{x}_{k^\ast}})^\top R(:,:,s)\vd^y_{\ell^y_{k^\ast}}d^z_{\ell^{z}_{k^\ast}}(s)$}
\ENDFOR
\ENDFOR
\STATE{\COMMENT{For clarity in the description only, we re-name here the residual and coefficients}}
\STATE{$\bar{\vR}= \vR;\; \bar{\vc}= \vc$}
\end{algorithmic}
\end{algorithm}
Due to computational complexity and memory requirements, 
 pursuit strategies using general dictionaries  
 can only be implemented on an image partitioned into small 
 blocks.
% Without loss of generality we assume that the 
%partition is realized  with 
%3D blocks all of the same size. 
We consider nonoverappling blocks. The approximation of 
 each block is carried out 
independently of the others. When the approximation of 
all the blocks is concluded, these are assembled 
together to produce the approximation of the whole image.
While the sparsity results yielded by the OMP3D and 
the SPMP3D methods are theoretically equivalent, 
we have seen that the latter implementation 
is much more economic in terms of storage demands. 
As discussed in Remark 2 below, this feature makes the SPMP3D
 algorithm suitable for 
possible GPU implementations using only the 
fast access shared memory. 
Assuming for simplicity in the notation 
that a 3D image is partitioned into cubes of size $\Nq^3$  
and the dictionaries  $\D_x$, $\D_y$ and $\D_z$ are 
all of the same size $\Nq \times r \Nq$, where $r>1$ is 
the redundancy of the 1D dictionary, the  
SPMP3D algorithm storage needs are as follows.
\begin{itemize}
\item[1.]
Two $\Nb^3$ arrays for the intensity block 
in the image partition and  the 
residual of the corresponding approximation.
\item[2.]
Three matrices of size $\Nq \times r \Nq$ for each dictionary, in 
case they are different.
\item[3.] A $r^2 \times \Nq^2$ array for the selection of 
indices in Algorithm~\ref{IP3D}.
\item[4.] A vector of $k$ real numbers to store the 
coefficients of the atomic decomposition and  
  $k$ vectors of size 3 to store the indices of the atoms in the atomic decomposition. 
 The value of $k$ 
is the total number of atoms in the approximation 
of the block. 
\end{itemize}
Since the stepwise complexity is dominated 
by the selection of indices (c.f. \eqref{select}),   
 within this setup it is 
 O($r^3 \Nq^5$) and for 
true color images O($r^3 \Nq^3$).

{{\em{Remark 2:}}
By considering blocks of size $8 \times 8 \times 8$ 
and dictionaries of redundancy $r=5$ in each dimension,
the above listed storage needs of the SPMP3D algorithm 
comfortably fit the fast access shared memory of 
a GPU in CUDA, which currently is 
48Kb. Indeed, in the worst-case scenario (corresponding to 
 an approximation of zero error using $k=8^3$ atoms 
 for the approximation of an $8 \times 8 \times 8$ block) 
 SPMP3D would require 38Kb to store most of 
 the arrays in double precision, except for those 
 with the selected indices which contain 
 integer numbers. This still leaves 10Kb for temporary 
 variables to be used within calculations.  
\subsection{Mixed Dictionaries}
\label{MD}
A key factor for the success in the
construction of sparse representations is
  to have a good dictionary.
 While a number of techniques for learning
dictionaries from training data have been proposed
in the literature
\cite{KMR03, AEB06, RZE10, TF11, ZGK11, HSK13, SNS15, 
WEB15},
 they are not designed
for learning large and highly coherent
 separable dictionaries.
Nevertheless,
 previous works \cite{RNBCP12, RNB13, LRN17,LRN18} have
demonstrated that
  highly  redundant and  highly coherent 
  separable dictionaries, which are easy to
construct, achieve remarkable levels of sparsity
in the representation of 2D images.
Such dictionaries are not specific to a particular class of
images. A discrimination is only made to take into
account whether the approximation is
carried out in the pixel intensity or in the wavelet domain.

As will be illustrated by the numerical examples in the 
next section, the approximation of the images we are 
considering are sparser when realized in the wavelet domain
(wd).
This entails the following steps:
\begin{itemize}
\item
Apply a wavelet transform to each channel $\vI_m, m=1,\ldots, \Nz$ to obtain the arrays $\vU_m, m=1,\ldots,\Nz$. 
For the numerical examples we have used the 
9/7 Cohen-Daubechies-Feauveau biorthogonal wavelet 
transform \cite{CDF92}.
\item
Approximate the array $\vU \in \R^{\Nxyz}$ exactly as
it is done in the pixel domain (pd).
\item
Apply the inverse wavelet transform to the
approximated planes to recover the approximated intensity channels.
\end{itemize}
The mixed dictionary we consider for the 2D approximation 
 consists of two sub-dictionaries:
A trigonometric dictionary, $\D_T^x$, which is the common sub-dictionary for the approximation in both domains, and a dictionary of
localized atoms, which contains atoms of different
shapes when used in each domain.

The trigonometric dictionary is the union of the
dictionaries $\Dc^x$ and $\Ds^x$ defined below:
\be
\begin{split}
\Dc^x&=\!\{w_c(n)
\cos{\frac{{\pi(2i-1)(n-1)}}{2\Mx}},i=1,\ldots,\Nx\}_{n=1}^{\Mx}\\
\Ds^x&=\!\{w_s(n)\sin{\frac{{\pi(2i-1)(n)}}{2\Mx}},i=1,\ldots,\Nx\}_{n=1}^{\Mx},\nonumber
\end{split}
\ee
where $w_c(n)$ and $w_s(n),\, n=1,\ldots,\Mx$
are normalization factors, and usually $\Mx=2\Nx$.
Thus, the trigonometric dictionary is
constructed  as $\D_T^x= \Dc^x \cup \Ds^x$.

For approximations in the pd
we add the dictionary,
$\DLp^x$, which is built by
 translation of the prototype atoms in the left graph of
 Fig.~\ref{proto}. This  type of dictionary is
 inspired by a general result holding for continuous spline
spaces. Namely, that spline spaces on a compact interval
can be spanned by dictionaries of B-splines
of broader support than the corresponding B-spline basis functions \cite{ARN05,RNX10}. Thus, the 
first 4 prototype atoms $\vh_i,\,i=1,\ldots,4$ in the left graph of
 Fig.~\ref{proto}  are generated by discretization of 
linear B-spline functions
of different support. For
 $m=1,2,3,4$ those functions are defined as follows:
\be
\label{B2}
h_m(x)=
\begin{cases}
\frac{x}{m} & \mbox{if}\quad 0 \leq x <  m\\
2-\frac{x}{m} & \mbox{if}\quad  m\leq x <2m \\
 0 &  \mbox{otherwise.}
\end{cases}
\ee
The remaining  prototypes, $\vh_5,\vh_6$ and $\vh_7$, in the left graph of Fig.~\ref{proto} 
are generated taken the derivatives of the 
previous functions:  $h_5(x)= (h_2(x))'$, 
$h_6(x)=(h_3(x))'$ and  $h_7=(h_4(x))'$. 
The corresponding dictionaries $\mathcal{D}_{H_m}, \, 
m=1,\ldots, 7$  are  built  
  by discretization of the variable
$x$ in \eqref{B2} and
 sequential translation of one sampling point, i.e.,
$$\D_{H_m}=\{w_{h_m}(n) h_m(i-n)|\Nx; i=1,\ldots,\Nx\}_{n=1}^{M},\quad
m=1,\ldots,7,$$
where the notation $h_m(i-n)|\Nx$ indicates the restriction to be
an array of size $\Nx$. The numbers $w_{h_m}(n), \,n=1,\ldots,M, \,m=1,\ldots,7$
are normalization factors. The dictionary $\DLp^x$ 
arises by the union of the dictionaries
 $\D_{H_m},\, m=1,\ldots,7$ i.e., 
$\DLp^x= \cup_{m=1}^7 \D_{H_m}$. The whole mixed dictionary 
$\Dpd^x$ is finally formed as $\Dpd^x= \Dc^x
\cup \Ds^x \cup \DLp^x$. For the other 
dimension we take $\Dpd^y=\Dpd^x$.
\begin{figure}[ht]
\begin{center}
%\vspace{-1cm}
\includegraphics[width=8cm]{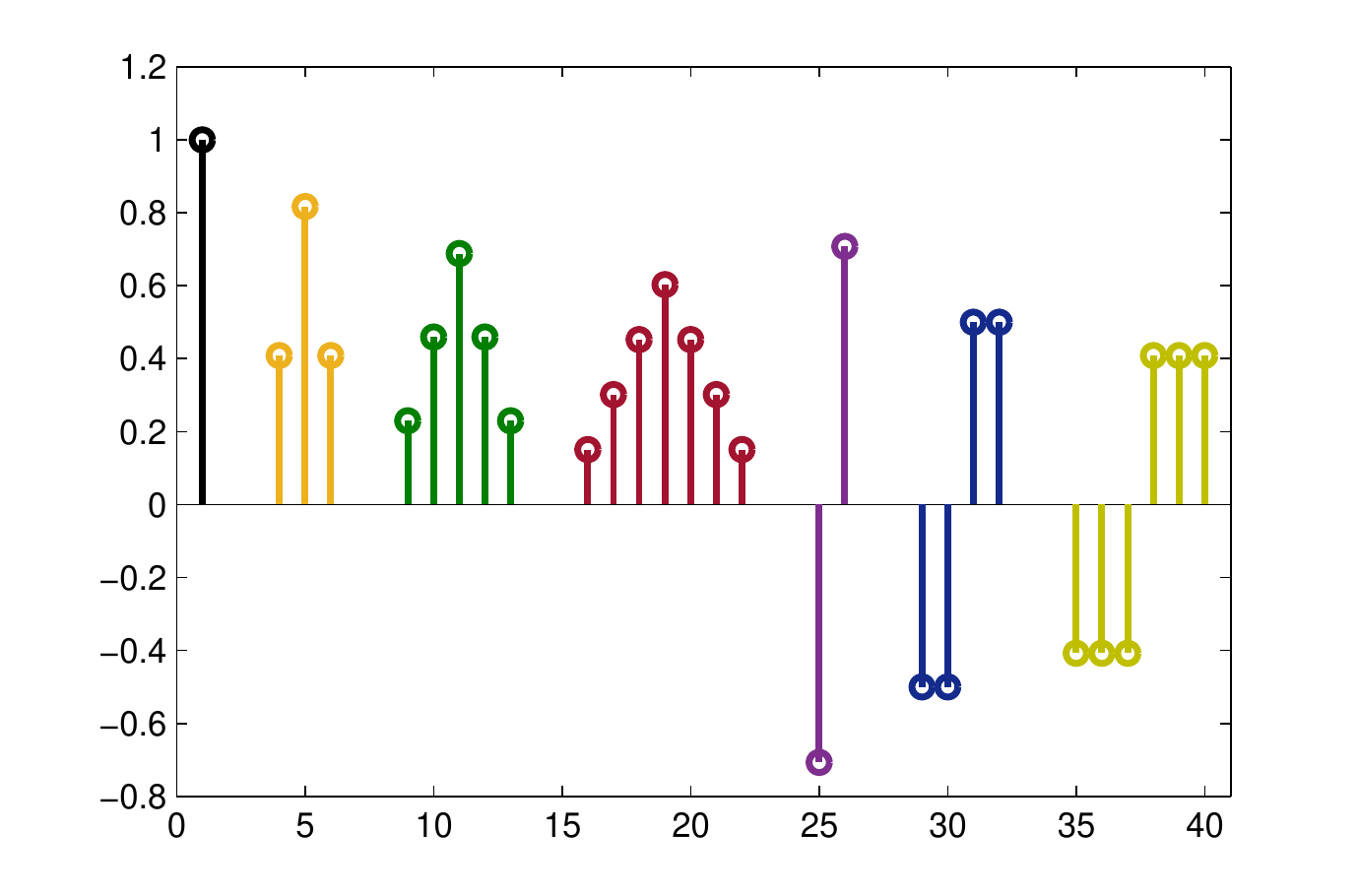}
\includegraphics[width=8cm]{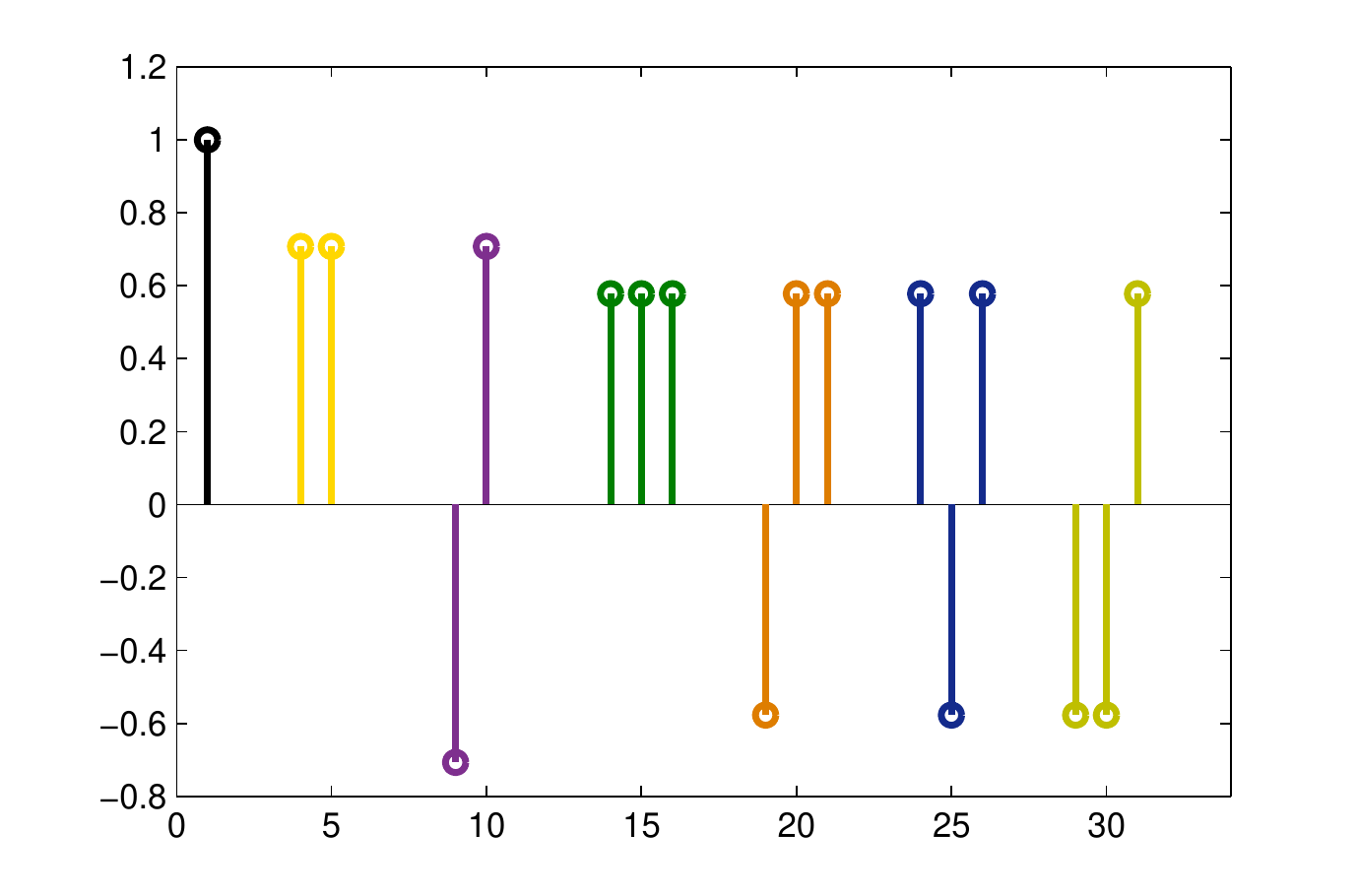}% \hspace{-0.1cm}
\end{center}
%\vspace{-1cm}
\caption{{\small{The left graph illustrates the 
prototype atoms which generate
  by translation the dictionaries $\D_{H_m}, \,
m=1,\ldots,7.$ The prototypes in the
 right graph  generate by translation the dictionaries 
$\D_{P_m},\, m=1,\ldots,7.$}}}
\label{proto}
\end{figure}

For approximations in the wd we use the
dictionary of localized atoms $\DLw^x$ 
 as proposed in \cite{LRN17},
 which is built by translation of
 the prototype atoms $\vp_i, \, i=1,\ldots,7$ 
in the right graph of
 Fig.~\ref{proto}. Notice that 
$\vp_1=\vh_1$ and  $\vp_3=\vh_5$. 
The remaining prototypes are given by the vectors:\\\\
$\vp_2=(1,1,0,0,\ldots,0)^\bot \in \R^{\Nx}, 
\vp_4=(1,1,1,0,\ldots,0)^\bot \in \R^{\Nx},
\vp_5=(-1,1,1,0,\ldots,0)^\bot \in \R^{\Nx},\\
\vp_6=(1,-1,1,0,\ldots,0)^\bot \in \R^{\Nx},
\vp_7=(-1,-1,1,0,\ldots,0)^\bot \in \R^{\Nx},$\\\\
The corresponding dictionaries $\mathcal{D}_{P_m}, 
\, m=1,\ldots,7$ are  built as in the previous case 
by sequential translation of one sampling point, 
$$\D_{P_m}=\{w_{p_m}(n) p_m(i-n)|\Nx; i=1,\ldots,\Nx\}_{n=1}^{M},\quad m=1,\ldots,7,$$
where the numbers $w_{p_m}(n), \,n=1,\ldots,M, \,m=1,\ldots,7$
are normalization factors.  The dictionaries $\D_{P_m}, 
m=1,\ldots,7$ give rise to $\DLw^x= \cup_{i=1}^7 \D_{P_m}$. 
The latter  generates the mixed dictionary 
$\Dwd^x= \Dc^x
\cup \Ds^x \cup \DLw^x$  and 
 $\Dwd^y=\Dwd^x$.

The corresponding 2D dictionaries 
$\Dpd= \Dpd^x \otimes \Dpd^y$ and
$\Dwd= \Dwd^x \otimes \Dwd^y$
are very large, but never used as such. All the
calculations are carried out using the 1D dictionaries.
In order to demonstrate the
gain in sparsity attained by the
approximation of 3D images by partitioning into 3D blocks, 
we use dictionaries $\Dwd$ and $\Dpd$ only for the approximation of the
single channel 2D images.
For the 3D case we maintain the redundancy of the 3D
dictionary equivalent to that of the 2D dictionary,
by considering the 1D dictionary
$\tDpd^x=\Dc^x
\cup \Ds^x \cup \D_{P_1}.$
Notice that $\D_{P_1}$ is the standard Euclidean basis for $\R^{\Nx}$,
also called the Dirac's basis,
i.e., the basis arising by translation of the first atom
in Fig.~\ref{proto}.
Notice that $\tDpd^x \subset \Dpd^x$ and  $\tDpd^x \subset 
\Dwd^x$. We also consider $\tDpd^y=\tDpd^x$ and
$\tDpd^z=\tDpd^x$, but taking $\Nx=\Nz$.
The redundancy of the  resulting dictionary
$\tDpd=\tDpd^x \otimes \tDpd^y \otimes \tDpd^z$
 is equivalent to the redundancy of the 2D dictionary
 $\Dpd$.  In 3D we use the same dictionary in both 
 domains 
$\tDwd= \tDpd$.  

\section{Numerical Results}
The merit of the simultaneous approximation of multiple 
channel images is illustrated in this section by recourse to 
two numerical examples. Firstly we 
   make the comparison between the sparsity 
produced by the joint approximation of the Red-Green-Blue 
(RGB) channel images partitioned into blocks of size 
$\Nb \times \Nb \times 3$  and 
 the sparsity obtained by the independent approximation of each channel partitioned into blocks of size $\Nb \times \Nb$. 
 Secondly, the full power of the approach is 
illustrated through the gain in sparsity attained by 
approximating hyper-spectral images partitioned into 
3D blocks, vs the plane by plane approximation.

In both cases, once the approximation 
of each 3D block $\vI_q$ in the
image partition is completed, for $q=1,\ldots,Q$ the
$\kq$-term
atomic decomposition of the corresponding block is
expressed in the form
\be
\label{atoq2r}
\vI_q^{\kq}= \sum_{n=1}^{\kq}
c_q(n) \vd^x_{\ell^{x,q}_n} \otimes \vd^y_{\ell^{y,q}_n} \otimes \vd^z_{\ell^{z,q}_n}.
\ee
The sparsity of the representation of an image of
dimension $N=\Nx \cdot \Ny \cdot \Nz$ is measured by the
Sparsity Ratio ($\SR$), which is defined as:
\be
\label{SR}
\text{SR}=\frac{N}{K},
\ee
where for the 3D representation $K=\sum_{q=1}^Q \kq,$
with $\kq$ the number of atoms in the
atomic decomposition \eqref{atoq2r}. For the
channel by channel decomposition of a $\Nz$-channel image,
each channel is partitioned into 
 $P=(\Nx \cdot \Ny)/\Nq^2$ blocks
$\vI_{p,z},\,p=1,\ldots,P$, which are approximated by 
the 2D atomic decompositions
\be
\label{atoq2c}
\vI_p^{\kpl}= \sum_{n=1}^{\kpl}
c_p^{p,l}(n) \vd^x_{\ell^{x,p,l}_n} \otimes \vd^y_{\ell^{y,p,l}_n} ,\quad l=1,\ldots,\Nz,
\ee
where the indices $\ell^{x,p,l}_n, \ell^{y,p,l}_n$ are 
selected for each channel $l$ by the OMP2D algorithm. 
Accordingly, the number $K$ in \eqref{SR} is given as
$K=\sum_{l=1}^{\Nz}\sum_{p=1}^{P}{\kpl}$, with 
$\kpl$ the number of atoms in the atomic decomposition
 \eqref{atoq2c}.

Notice that the $\SR$ is a measure of the
reduction of dimensionality for representing an image.
The larger the value of the SR the smaller the
dimensionality of the atomic decomposition representing
the whole image.
The required quality of the  
approximation is ensured  with respect to  the
Mean Structural SIMilarity (MSSIM) index~\cite{ssim,ssimex}
 and the classical Peak Signal-to-Noise Ratio ($\PSNR$),
which for a 3D image is defined as
\be
\label{psnr}
\PSNR=10 \log_{10}\left(\frac{{(\IM)}^2}{\MSE}\right),
\quad
\MSE=\frac{\|\vI - \vI^K\|_{\F}}{\Nx \cdot \Ny \cdot\Nz},\nonumber
\ee
where $\IM$ is the maximum intensity range and
$\vI^K$ the image approximation.
\subsection{Example I} 
In this example we use the Kodak data set consisting 
of 24 true color images shown in Fig.~\ref{Kodak}. 

\begin{figure}[H]
\begin{center}
\includegraphics[width=14cm]{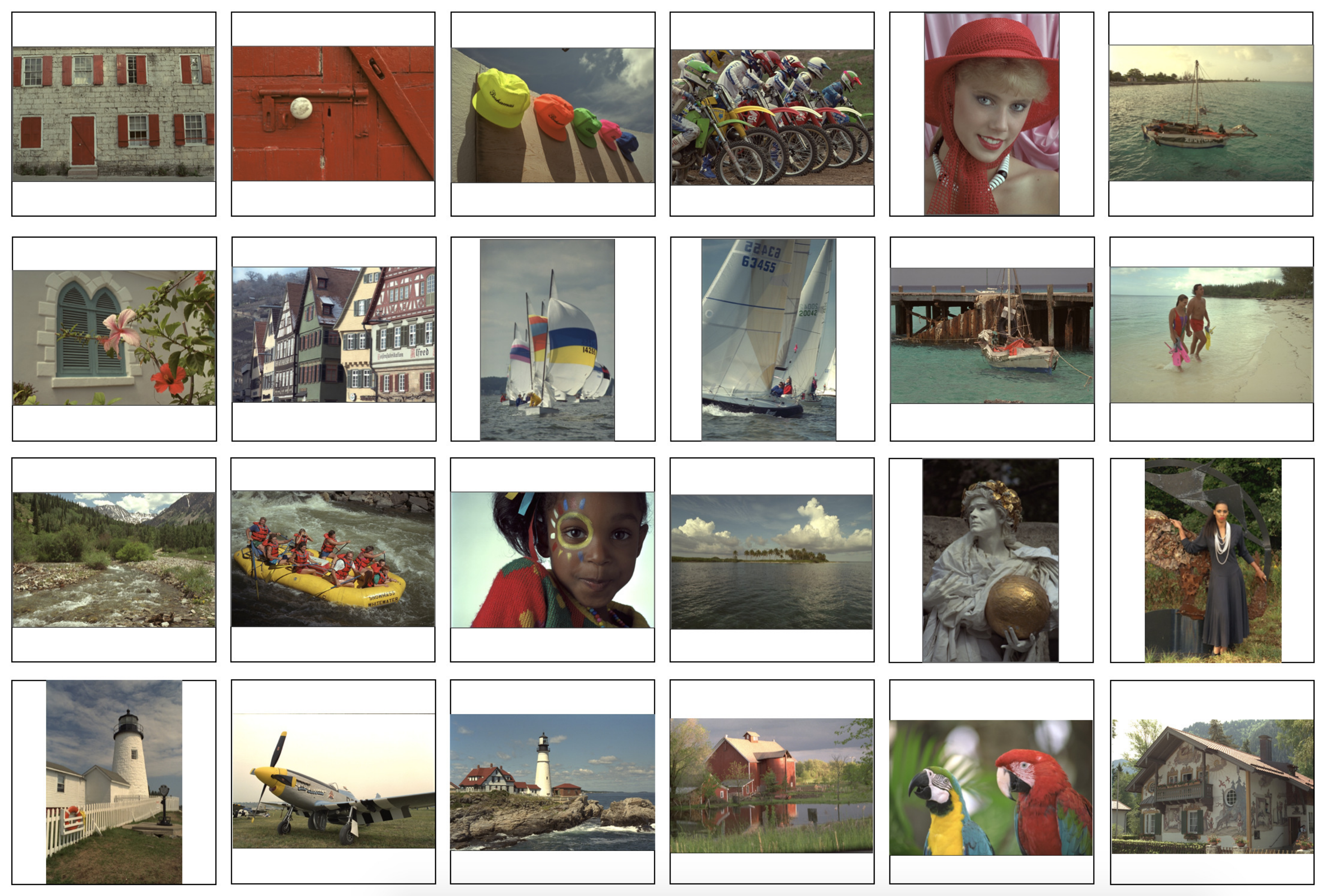}
\end{center}
\caption{{\small{Illustration of the
Kodak data set consisting of 24 true color images,
credit  Rich Franzen \cite{kodak}.
The  size of these
 images is $768 \times 512 \times 3$, for most of them,
except for numbers
4, 9, 10, 17, 18 and 19, which are of size
$512 \times 768 \times 3$.}}}
\label{Kodak}
\end{figure}

The approximations are realized in both domains 
 by  maintaining the same redundancy in the 
 2D and 3D dictionaries.
For the independent approximation of the 2D channels
the partitions are realized with blocks of  size
$8 \times 8$ and
$16 \times 16$ (a partition of block size $24 \times 24$ 
does not improve results for this data set). 
Accordingly, the simultaneous
approximation of the 3 color channels involves partitions of
block size $8 \times 8 \times 3$ and
$16 \times 16 \times 3$ respectively.

As already discussed, for the independent approximation 
of the 2D channels
 we consider the dictionaries  
$\Dpd$ (in the pd) and $\Dwd$ (in the wd) 
as given in Sec.~\ref{MD}. For the simultaneous 
approximation of the 3 channels we 
consider the dictionaries $\tDpd$  
given in the same section. Both dictionaries have
redundancy of 125.

The average values of SR ($\SRm$),
with respect to the 24 images in the set,
are given in Table ~\ref{TABLE1} for the 
approaches and partitions indicated by the first column. 

\begin{table}[H]
\begin{center}
\begin{tabular}{|l||r|r|| r| r||}
\hline
$\PSNR$ & \multicolumn{2}{|c||} {45 dB} & \multicolumn{2}{|c||}{41 dB}\\ \hline \hline
 & $\SRm$ & std&  $\SRm$&  std\\ \hline  \hline
pd 2D $8\times 8$& 6.2 & 2.0& 9.1 & 3.5 \\ \hline
pd 3D $8\times 8 \times 3$& 10.3& 2.9& 16.1& 5.5 \\ \hline
%pd 3D $8\times 8 \times 3$\,(b)& 11.7& 3.2& 17.1& 5.5\\ \hline
wd 2D $8\times 8$& 7.1&  2.6& 11.8& 5.8 \\ \hline
wd 3D $8\times 8 \times 3$& 11.6&  3.8& 20.9& 9.2 \\ \hline
\hline
%wd 3D $8\times 8 \times 3$\,(b) & 15.4&5.1& 25.5 & 11.3 \\ \hline \hline
pd 2D $16\times 16$& 7.1 & 2.5 &11.1 & 5.0 \\ \hline
pd 3D $16\times 16\times 3$& 11.6& 3.6 & 18.8& 7.5 \\ \hline
%pd 3D $16\times 16\times 3$\,(b)& 13.7& 4.2 & 23.6 & 10.2 \\ \hline
wd 2D $16\times 16$& 7.5 & 2.7 & 12.0& 6.2 \\ \hline
wd 3D $16\times 16\times 3$& 12.4& 3.9 & 20.4 & 8.9 \\ \hline 
%wd 3D $16\times 16\times 3$\,(b)& 15.9& 5.5 & 26.6&12.8 \\ \hline \hline
Thresholding in the wd &3.2& 1.1 & 4.9 & 2.6 \\ \hline \hline

\end{tabular}
\caption{{\small{Mean value of the SR, with respect to the 24 images in the set, obtained with the 2D and 3D approximations in
both the $\mathrm{pd}$ and $\mathrm{wd}$
 for two different sizes of the image partition.
The last row in the table gives the results
corresponding to standard nonlinear thresholding of
 wavelet coefficients, to achieve the same quality of the
approximation as with
the dictionaries: $\PSNR=45{\mathrm{dB}}$ (left half)
and $\PSNR=41{\mathrm{dB}}$ (right half).}}}
\label{TABLE1}
\end{center}
\end{table}

All the results in the left half of the table  
correspond to $\PSNR =45$ dB and all the results in the 
right half correspond to $\PSNR =41$ dB.
The third and fifth columns give the 
standard deviations (std).
 For completeness we have also produced the
$\SRm$ rendered by nonlinear thresholding of the wavelets
coefficients (last row in the table). 
Notice that the
 resulting sparsity is   
 poor in comparison with the other 2D results.

All the results were obtained in the MATLAB environment 
 on a notebook 2.9GHz dual core i7 3520M CPU and 4GB 
of memory. 
For the channel by channel approximation  
a C++ MEX file implementing OMP2D was used. 
For the 3D approximation SPMP3D was implemented by a
C++ MEX file.

As observed in Table~\ref{TABLE1} the largest 
$\SRm$ is achieved    
 in the wd and partition $16 \times 16 \times 3$ 
(c.f. last but one row of Table~\ref{TABLE1}). However, 
the results obtained by 
partition $8 \times 8 \times 3$ are very close 
(c.f. last row of the upper half of 
 Table~\ref{TABLE1}) and 
 constitute a better tradeoff between SR and approximation
 time. 

 Fig.~\ref{SRswd} shows the actual values of
 SRs for this partition in the wd for each of the 24
 images in the data set (c.f. Fig.~\ref{Kodak}).
The average time for the
 3D approximation was 53 s per image.

\begin{figure}[htp!]
\begin{center}
\includegraphics[width=12cm]{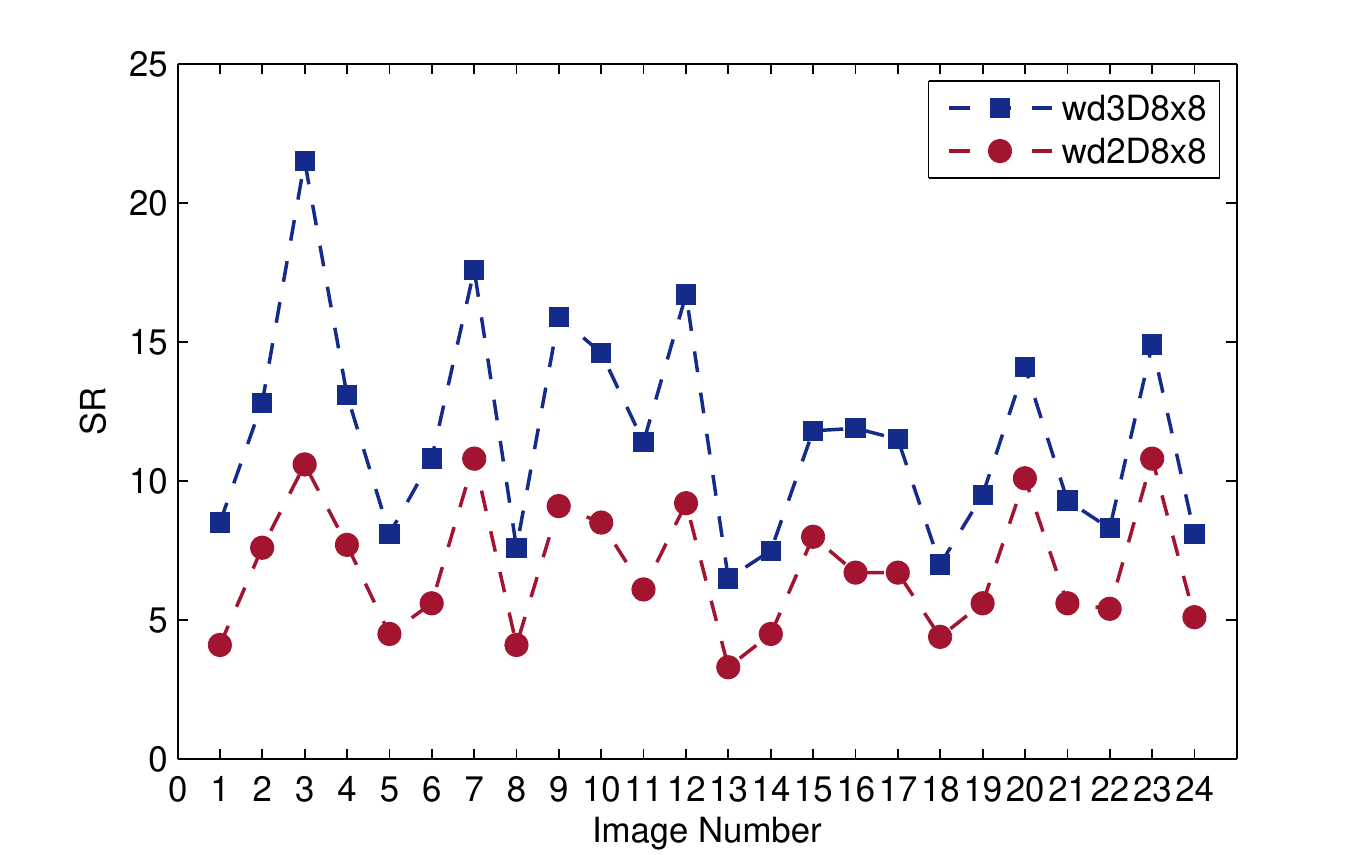}
\end{center}
\caption{{\small{SR for the 45dB
approximation, in the wd, of each of the 24 images
in the Kodak data set (c.f. Fig.~\ref{Kodak} enumerated from top left to bottom right).
The results for the independent approximation of
 each 2D color channel are represented by
the filled circles
 and those corresponding to the
simultaneous
approximation of the 3 channels are represented by the
filled squares. The corresponding
partitions are of size $8\times8$ and $8 \times 8 \times 3$.}}}
\label{SRswd}
\end{figure}

\begin{figure}[htp!]
\begin{center}
{\bf{3D\,\, SR=63.5\,\, PSNR=38.4\;\;\;\;\;\;\;\;\;\;\;\;\;\;\;\;\;\;\;\;\;2D\,\,SR=63.5\,\, PSNR=24.8}}

\includegraphics[width=8cm]{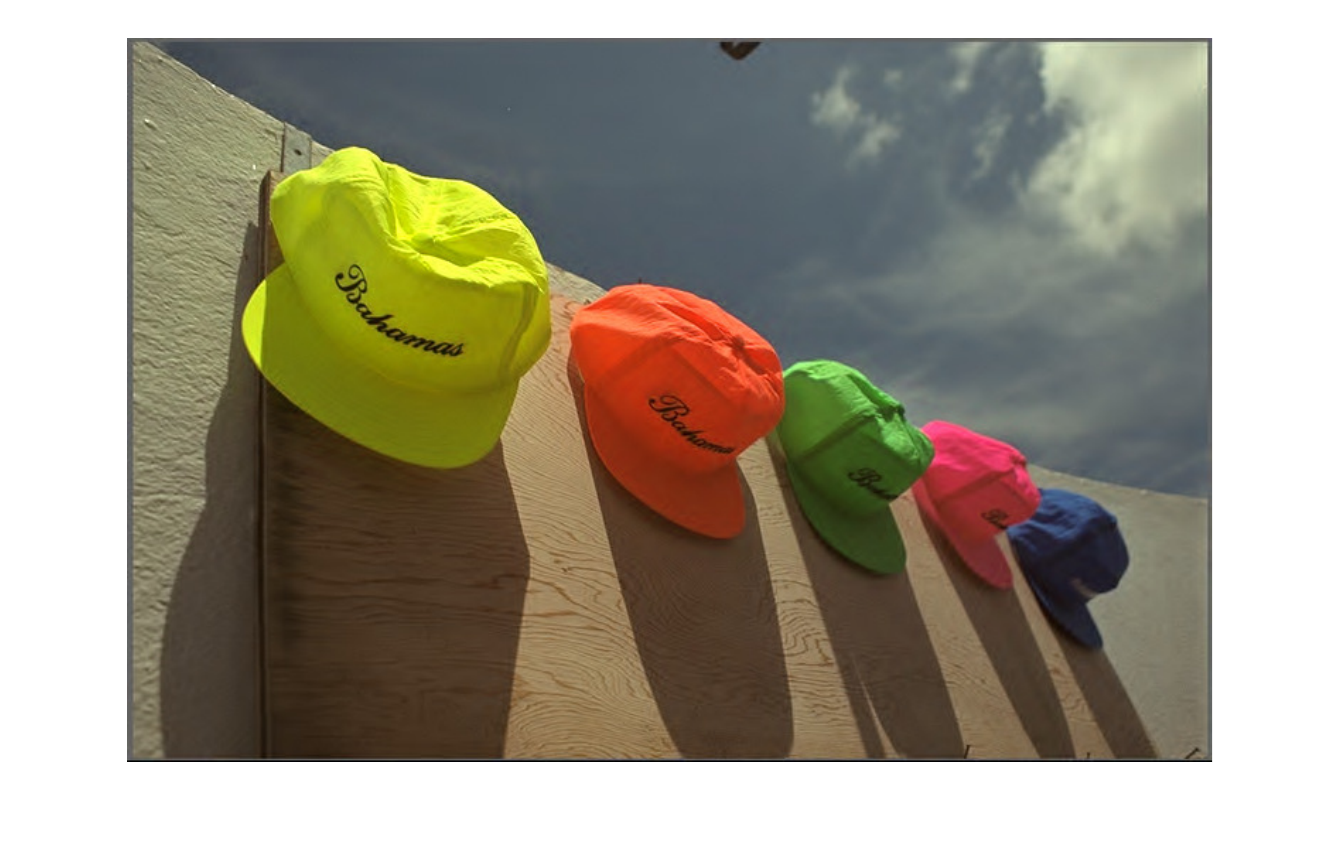}
\includegraphics[width=8cm]{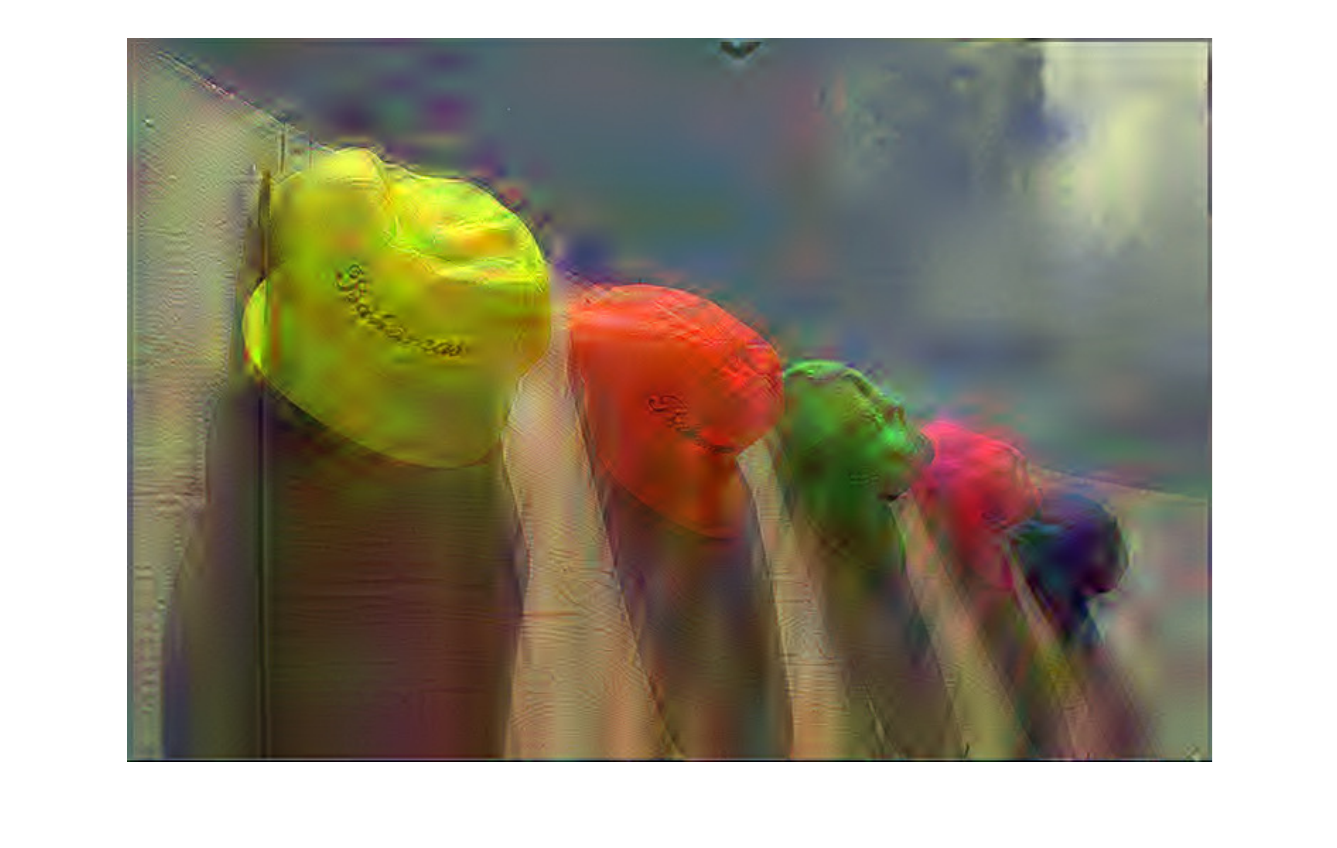}\\
{\bf{3D\,\,SR=63.5; PSNR=35.9\;\;\;\;\;\;\;\;\;\;\;\;\;\;\;\;\;\;\;\;\;2D\,\,SR=63.5; PSNR=22.1}}

\includegraphics[width=8cm]{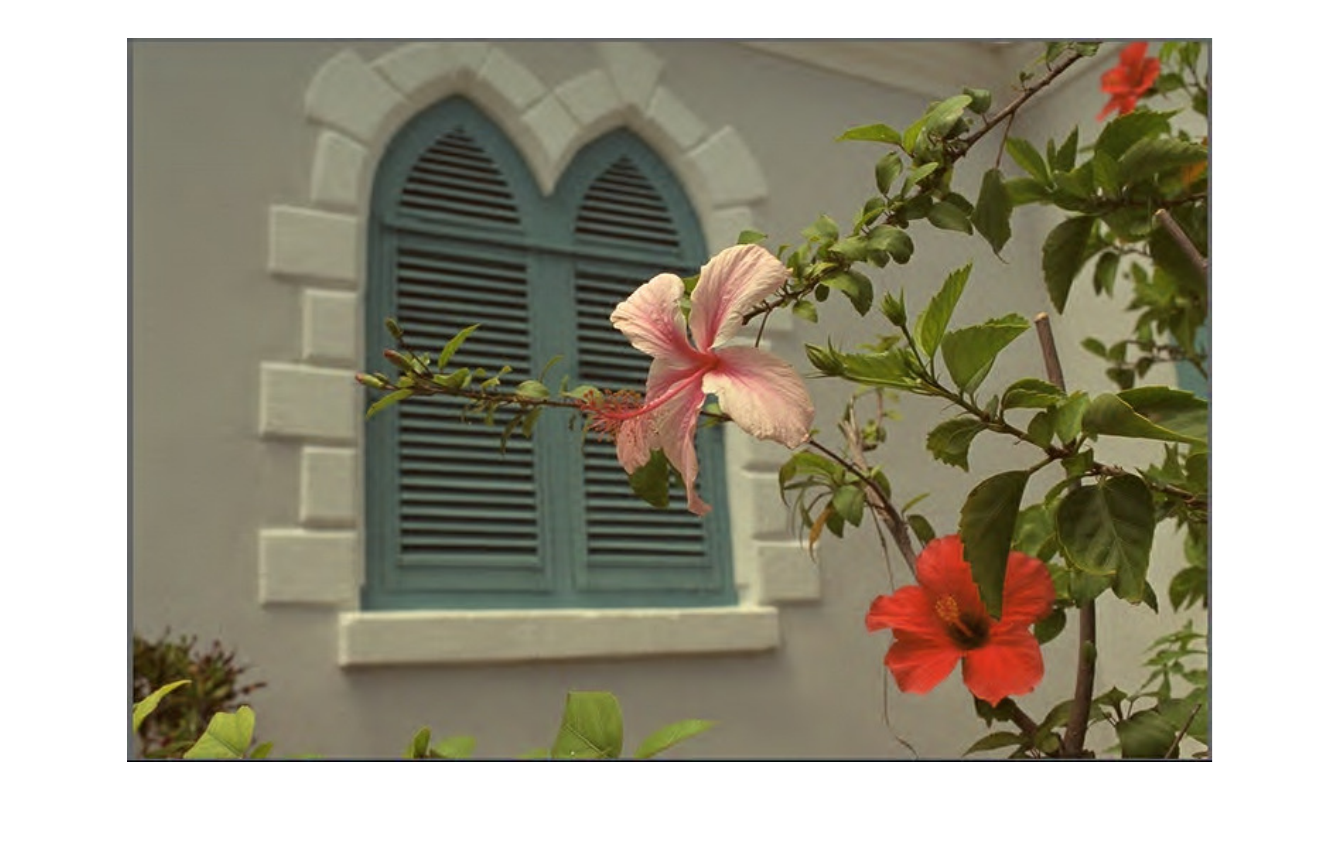}
\includegraphics[width=8cm]{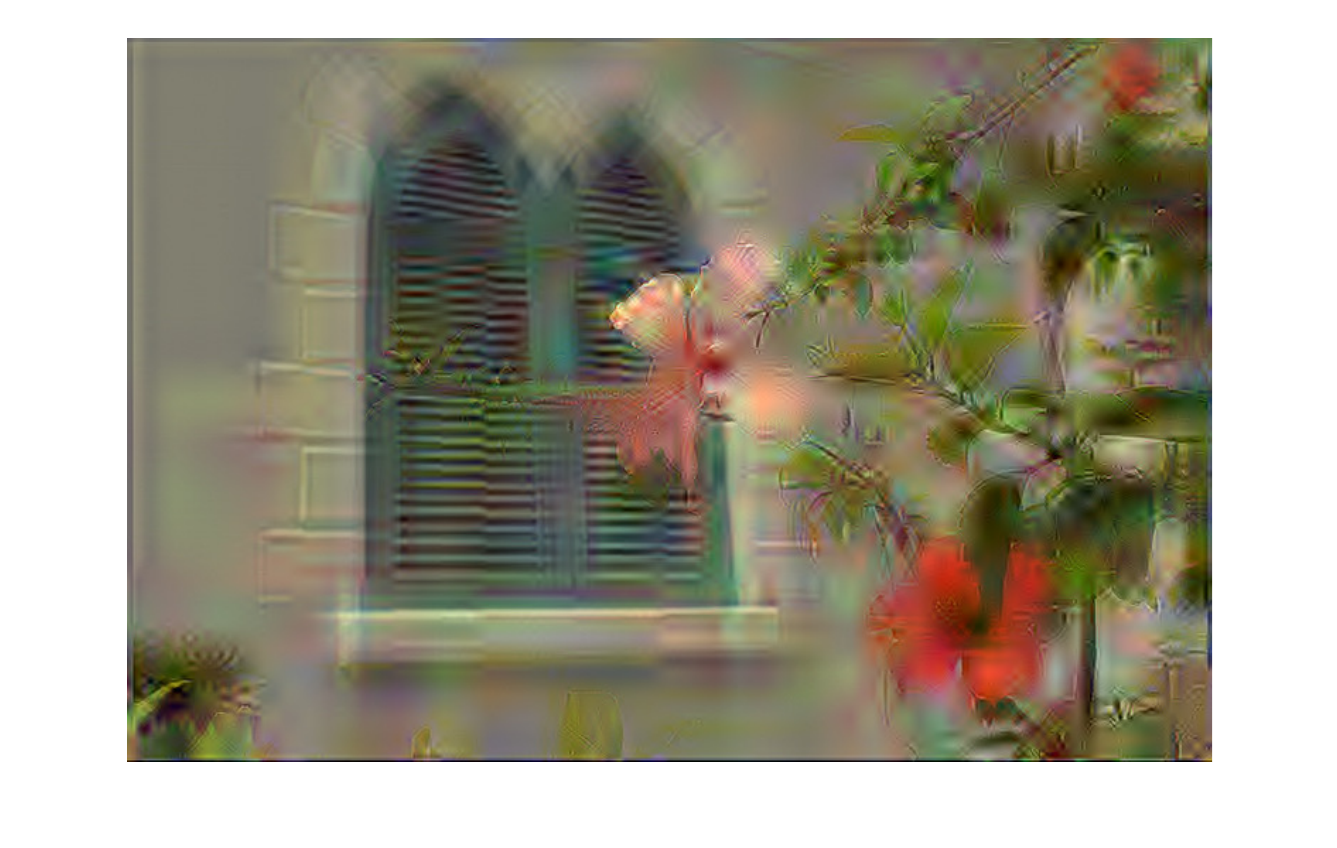}\\
{\bf{3D\,\,SR=63.5; PSNR=37.6\;\;\;\;\;\;\;\;\;\;\;\;\;\;\;\;\;\;\;\;\;2D\,\,SR=63.5; PSNR=24.9}}

\includegraphics[width=8cm]{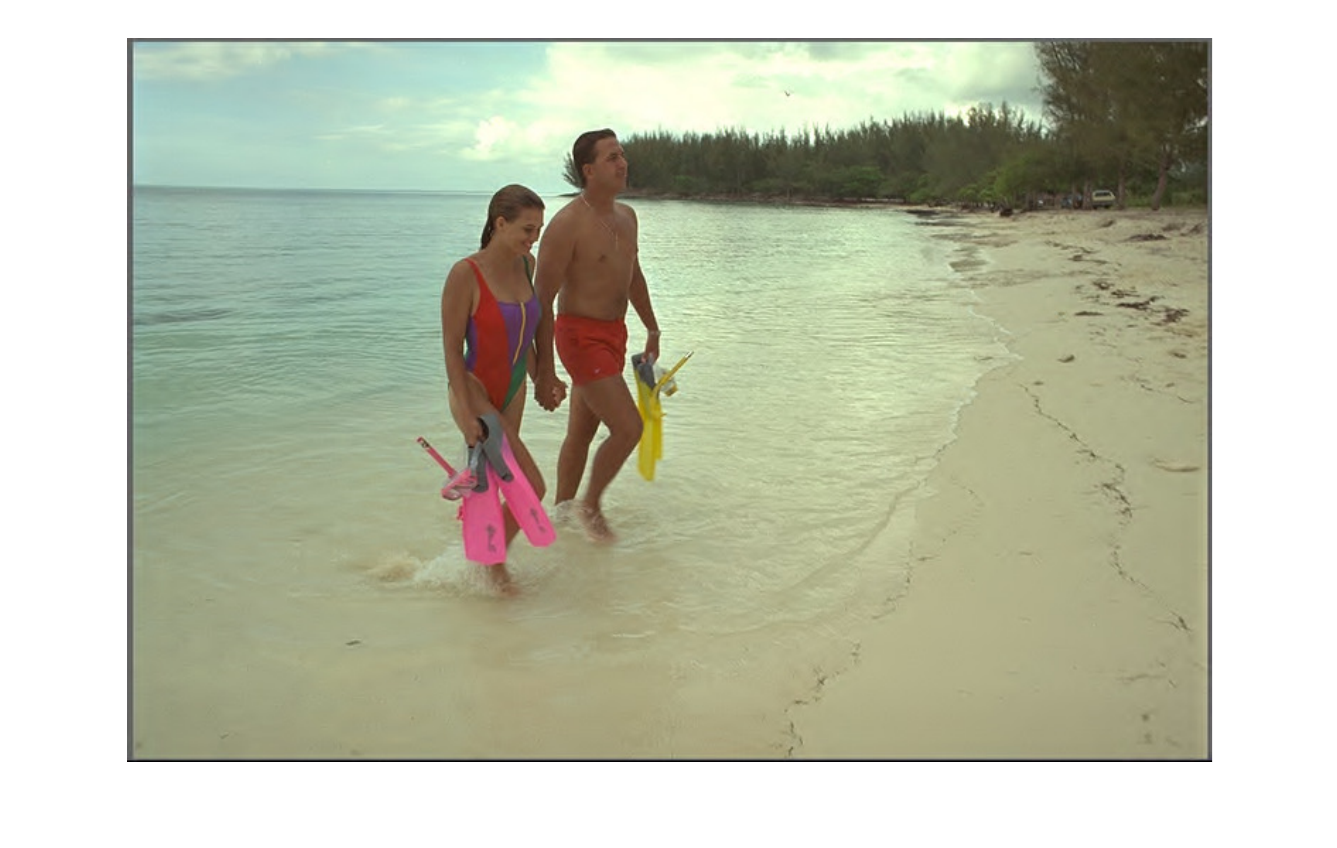}
\includegraphics[width=8cm]{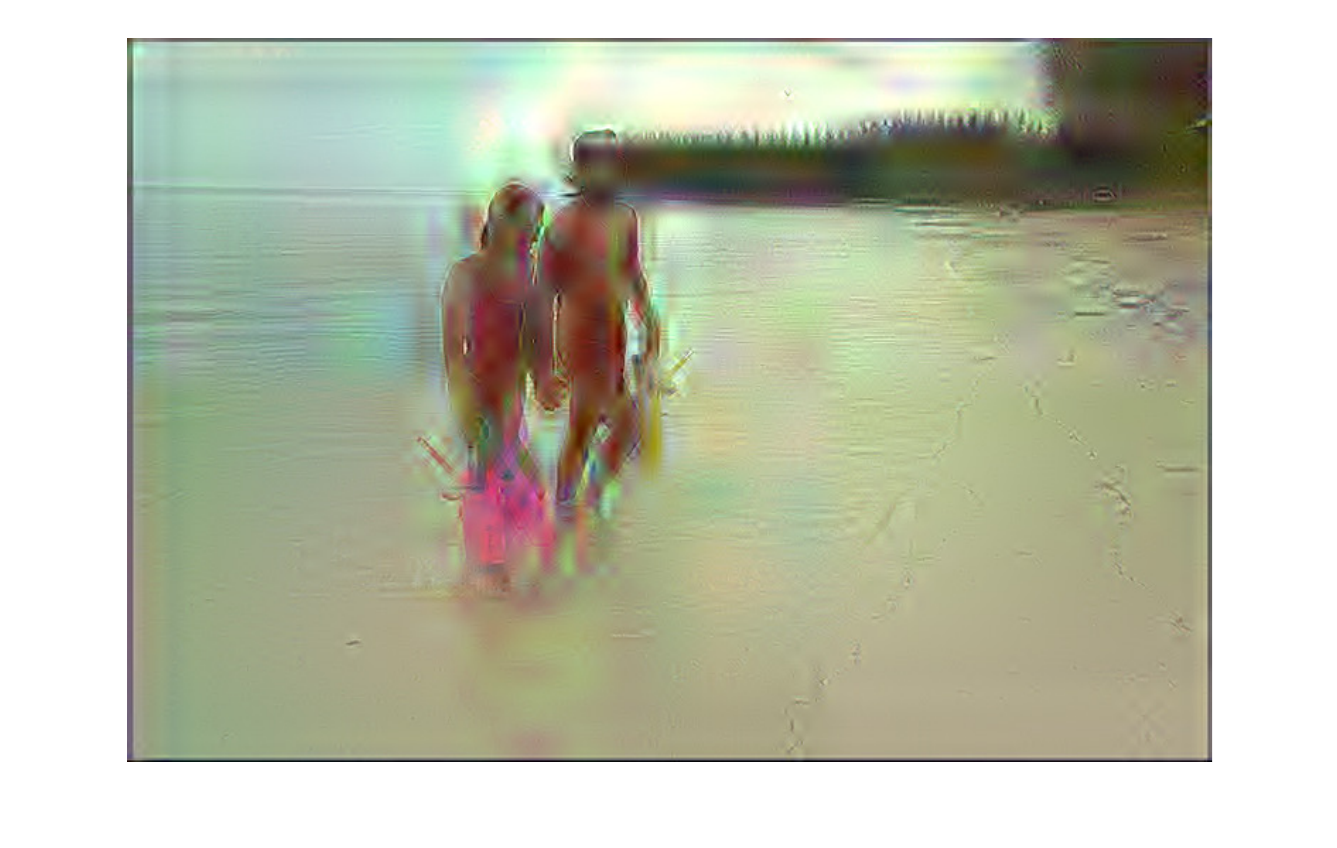}\\
\end{center}
\caption{{\small{Approximations of Images 3, 7 and 12 in the 
Kodak data set, for SR=63.5. The images on the left 
are the 3D approximations. The images on the right 
are the 2D channel by channel approximations.}}}
\label{appro}
\end{figure}

Fig.~\ref{appro} demonstrates the gain in visual
quality  obtained  when the approximation of Images
3, 7 and 12 are
realized simultaneously in 3D, instead of
independently for each 2D channel. In both cases the SR is
fixed at a high value SR=63.5. While the
 3D approximation is still of good quality
 (c.f. images on the left in Fig.~\ref{appro}) the
distortion of the channel by channel approximation
is very noticeable even at the scale of the figure
(c.f. images on the right in Fig.~\ref{appro}).

As a final remark 
it is worth noting that the number $\kq$ of atoms in the 
 approximation of each block $q$ of an image partition 
 produces a meaningful summary of local sparsity.

\begin{figure}[H]
\begin{center}
\includegraphics[width=7cm]{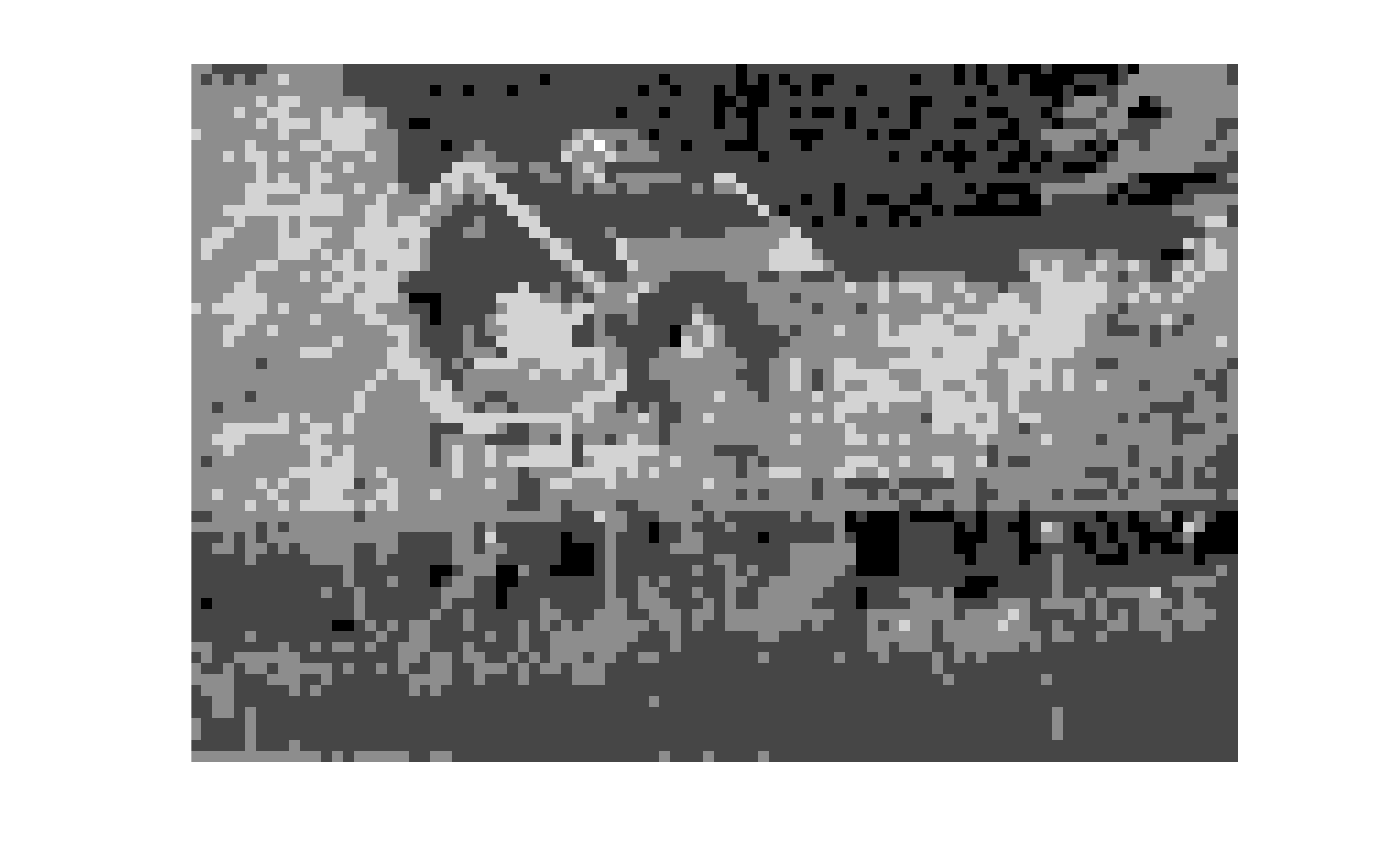}
\includegraphics[width=7cm]{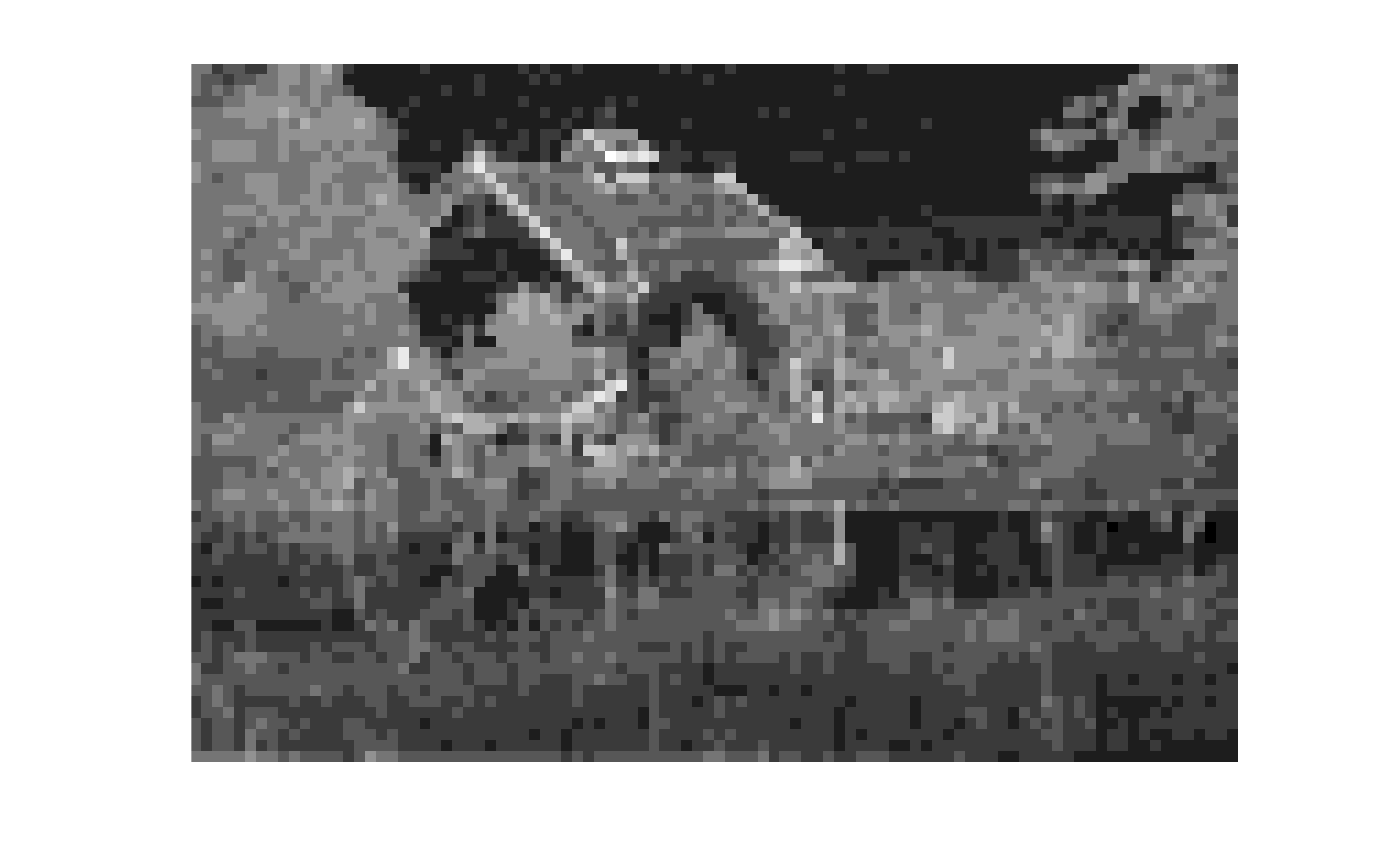}
\includegraphics[width=7cm]{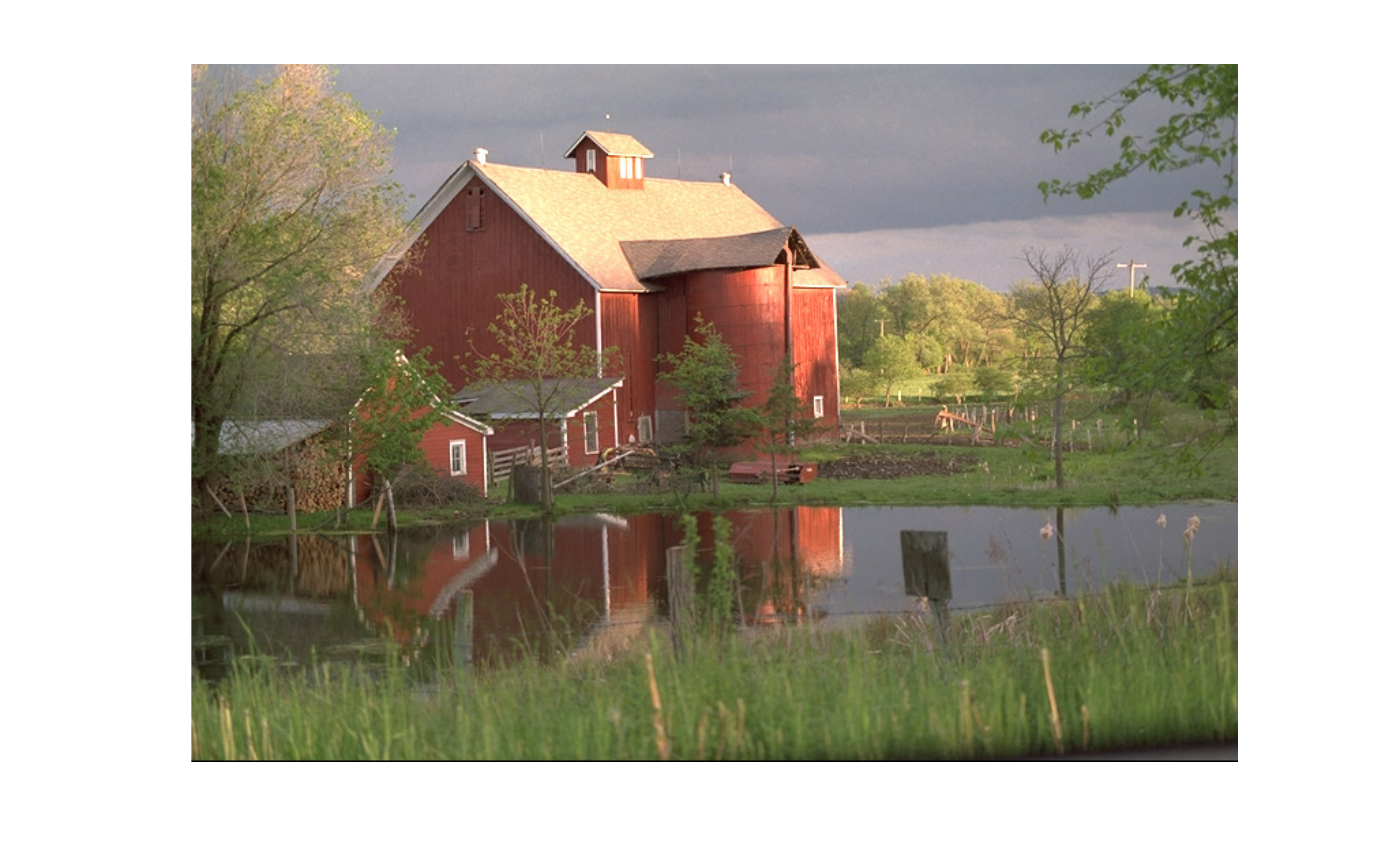}
\end{center}
\caption{{\small{The upper graphs are a
 representation of the piecewise sparsity corresponding
 to  Image 22 in the Kodak data set.
 Both graphs are arrays of $64 \times 96$ points.
 Each point corresponds to the number $k_q$ of
 atoms in the approximation
 of a block $q$.
 The left graph corresponds
 to the 2D approximation and the right graph to the
 3D approximation. The lower graph is the
 image given as 3 channels of $512 \times 768$
 pixels each.}}}
\label{summary}
\end{figure}

 The upper graphs of Fig.~\ref{summary} 
 are a representation of the piecewise sparsity 
 corresponding to Image 22 in the Kodak data set.
 Both graphs are arrays of $64 \times 96$ points.
 Each point corresponds to the number $\kq$ of
 atoms in the approximation of a block $q$.
 The left graph corresponds
 to block size $8 \times 8$ in the 2D approximation, 
 by taking the average 
 $\kq$ over the three channels in the block, 
 which is roughly the 
  $\kq$-value corresponding to the equivalent block in the 
  gray scale image.
  The right graph corresponds to $\kq$ for each block 
  of size $8 \times 8 \times 3$ in the 3D approximation. Both approximations 
 are realized in the pd. 
 The lower graph is the 
 image given as 3 channels of $512 \times 768$
 pixels each. It follows from the figure that the 
  points corresponding to the 3D approximation 
  give mode details about the image. 

\subsection{Example II}
We consider now the approximation of the 
hyper-spectral images illustrated in Fig.~\ref{hsi}. 
Details on the
images acquisition and processing
are described in \cite{FNA04, FAN06, NAF16}.

\begin{figure}[ht]
\centering
%\vspace{-0.9cm}
\includegraphics[scale=0.5]{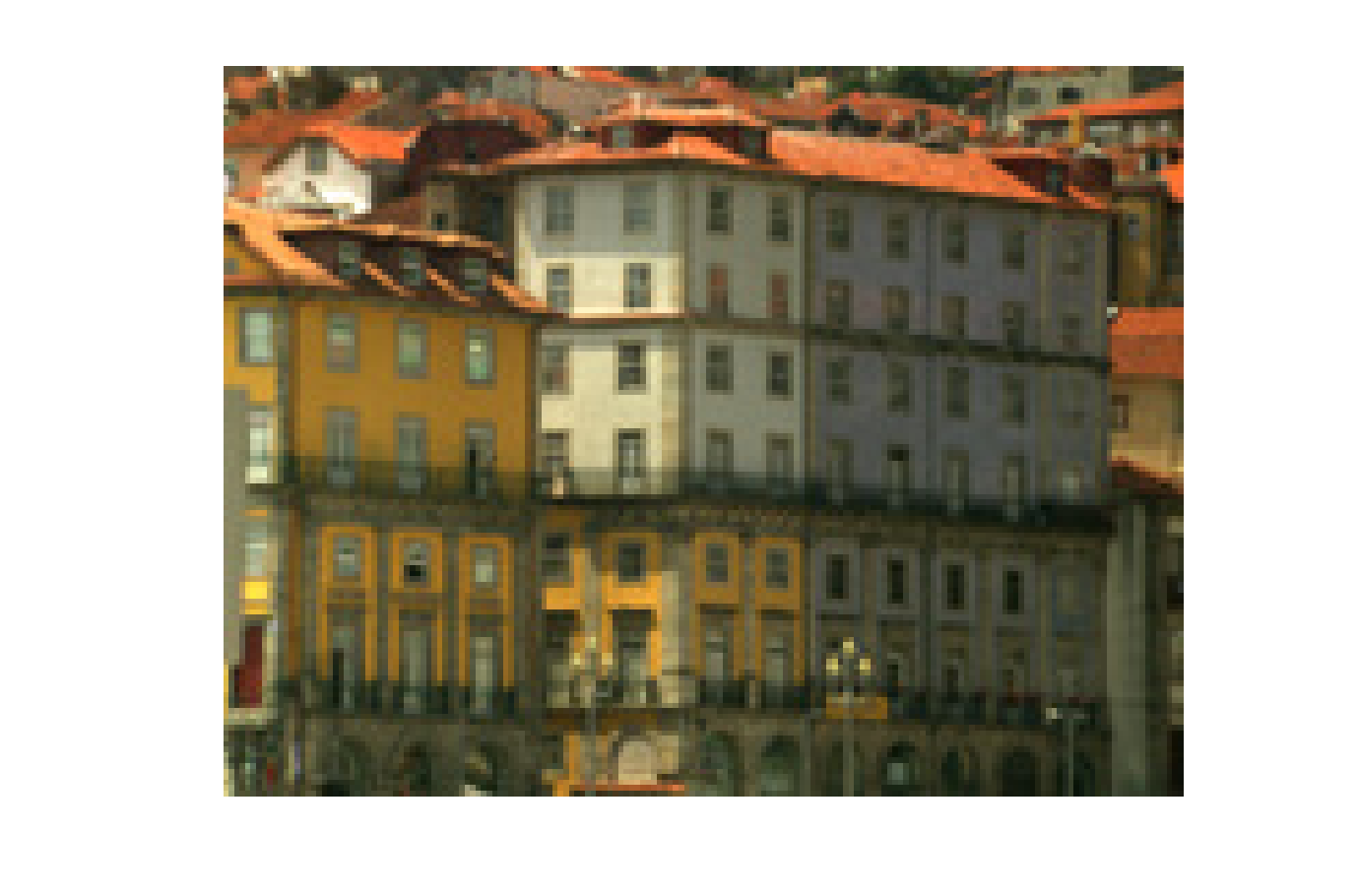} \hspace{-0.2cm}
\includegraphics[scale=0.5]{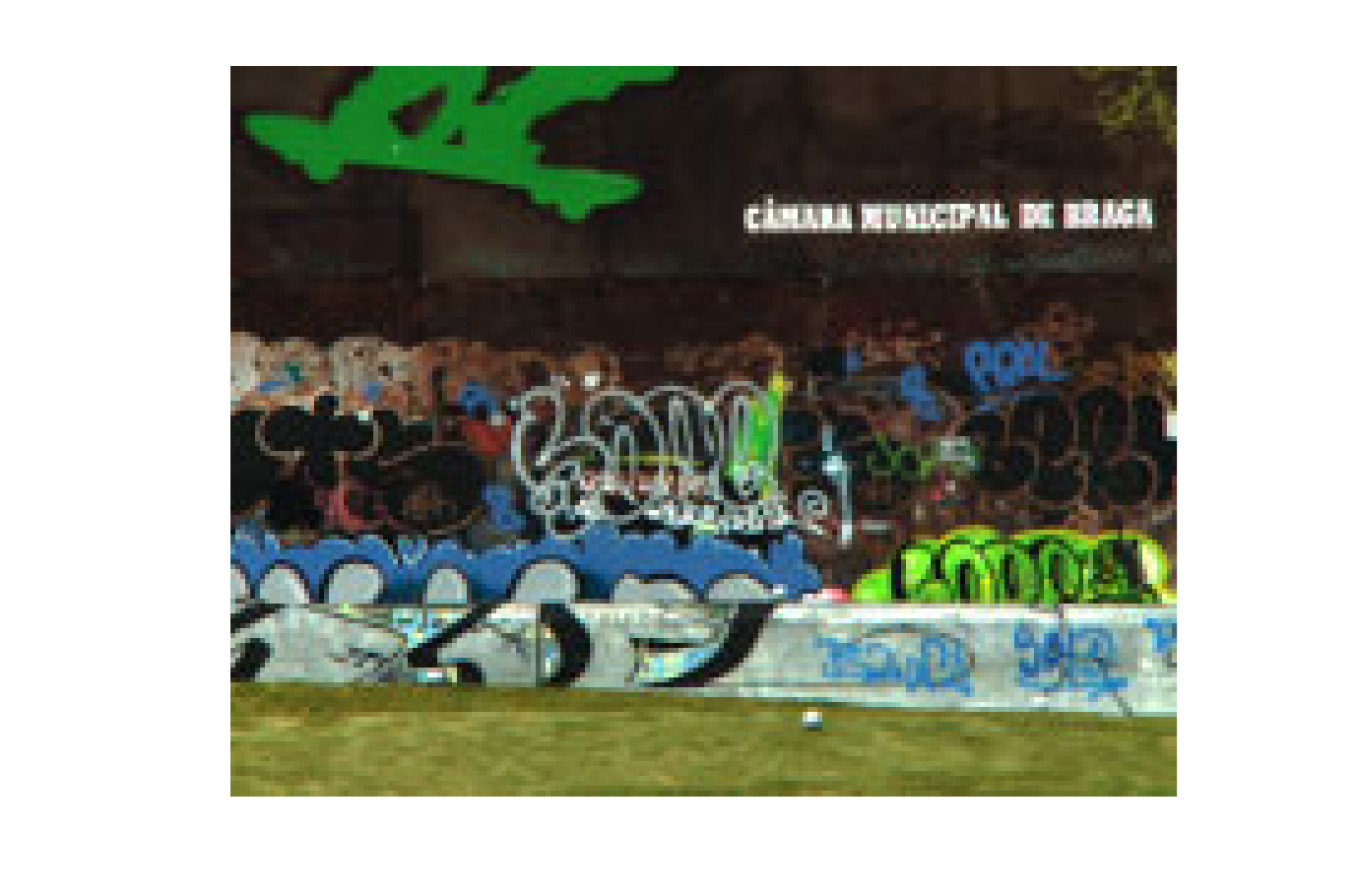}\hspace{-0.2cm}\\
%\hspace*{0.1cm}
\includegraphics[scale=0.5]{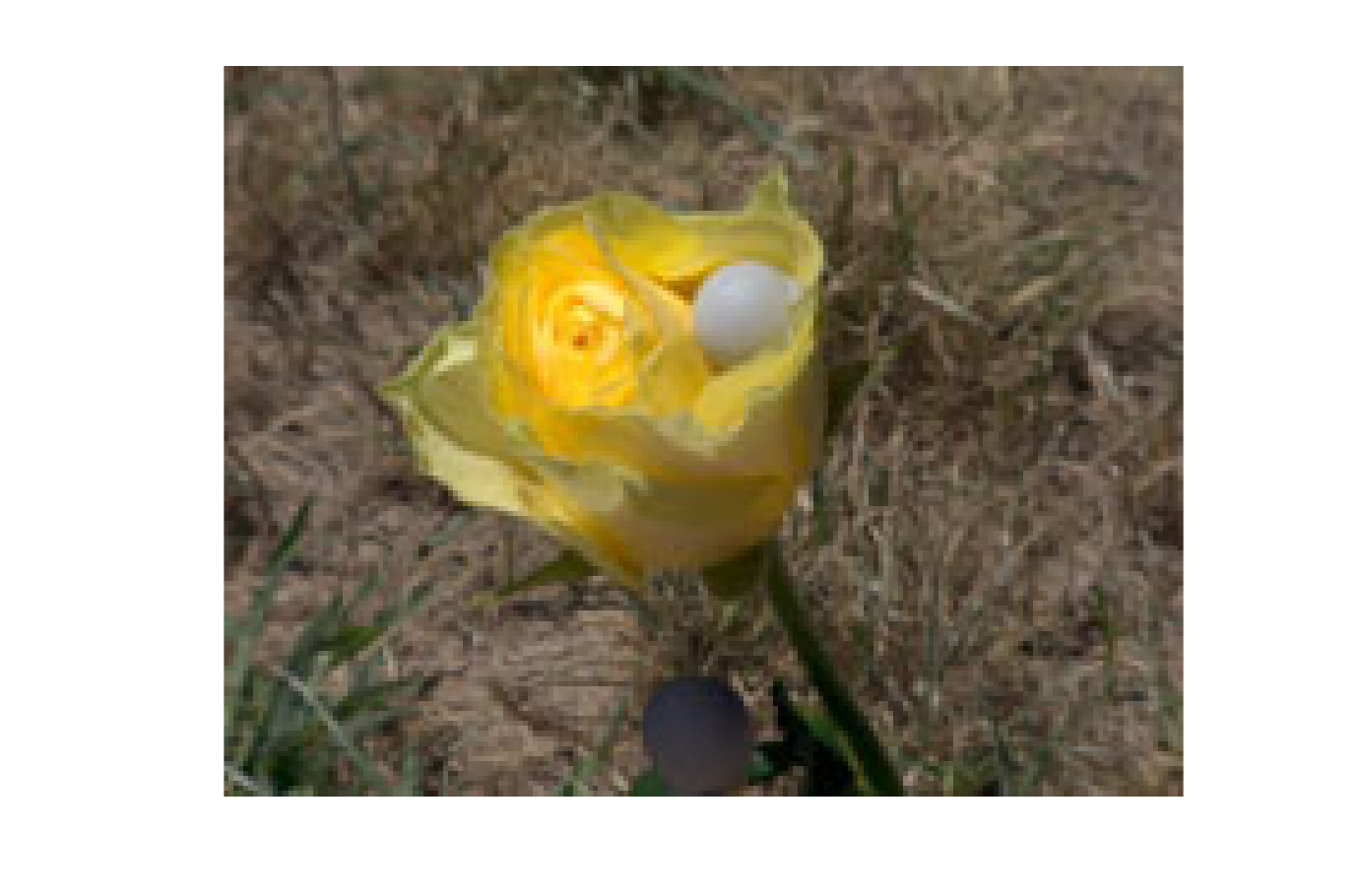} \hspace{-0.2cm}
\includegraphics[scale=0.5]{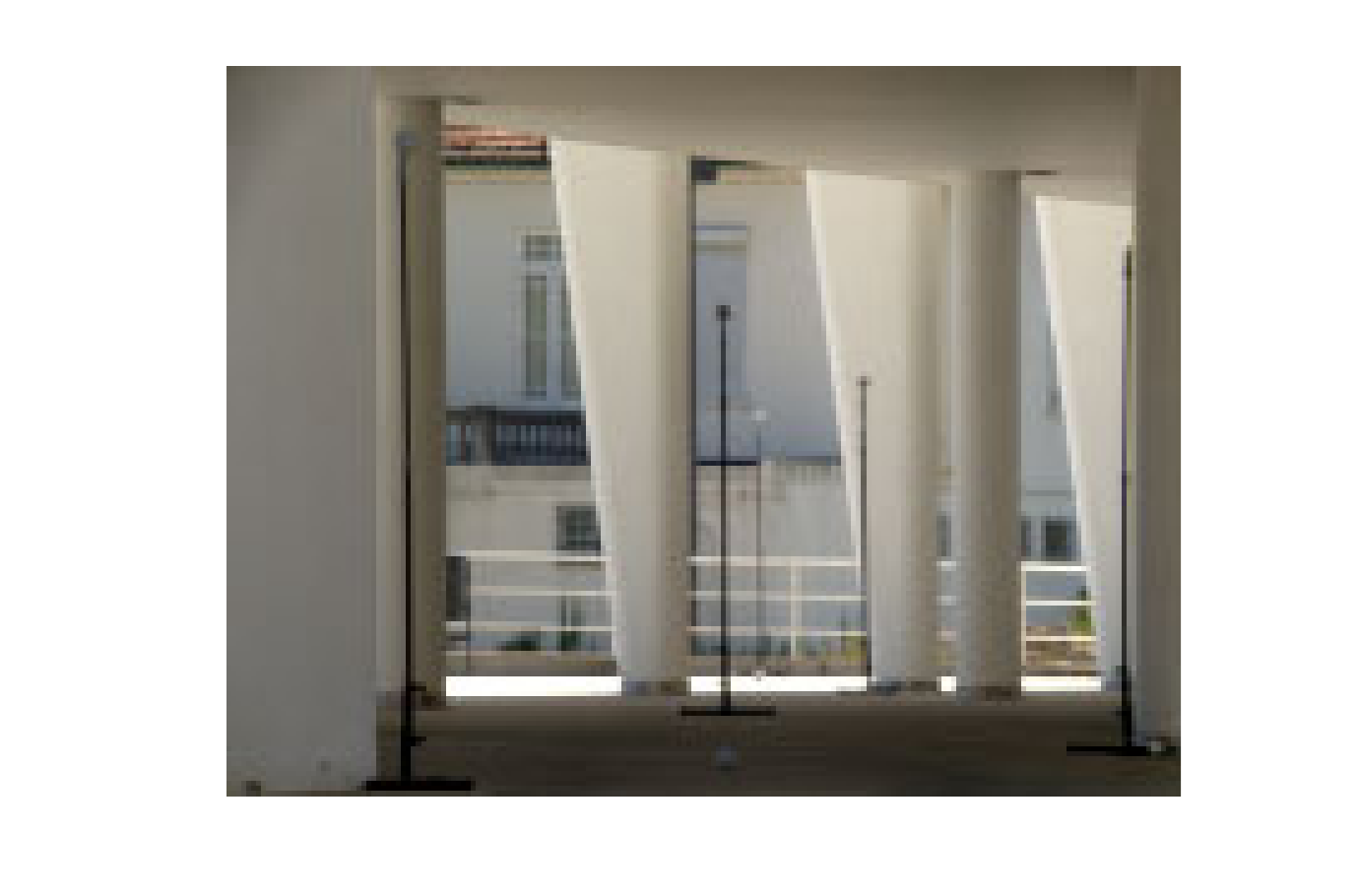} %\hspace{-0.5cm}
\caption{{\small{Illustration of the hyper-spectral images
 available on \cite{HSI04} and \cite{HSI}. From top left to bottom right
in Table.~\ref{TABLE3} are labelled as Ribei.,  Graff., Rose, and Col. The of size of all four images is
$1016 \times 1336\times 32$ pixels.}}}
\label{hsi}
\end{figure}

All four images are of size 
$1016 \times 1336\times 32$,
and have been approximated 
in partitions of block size  
  $\Nb \times \Nb$, with
$\Nb=8,16$, and 24 
for the 2D approximation, and 
  $8 \times 8 \times 8$ for the 3D approximation.
For the 2D channel  by channel approximation we use the 
dictionaries $\Dpd$  and $\Dwd$ as defined 
in Sec.~\ref{MD}. 
For the 3D approximation we maintain the  
redundancy as in 2D using the dictionary  
$\tDpd$  introduced
 Sec.~\ref{MD}  and  $\tDwd^a=\tDpd^a$.

Because the range of intensity varies across the images, 
 in order to compare SRs  with different approaches we fix the  
Signal to Noise Ratio ($\SNR$)
\be
\label{snr}
\SNR=10\log_{10}\left(\frac{\| \vI\|^2_{\F}}{\|\vI - \vI^K\|^2_{\F}}\right).
\ee
\begin{table}[H]
\begin{center}
\begin{tabular}{|l||c|c|c| c||}
\hline
Image & Ribei. & Graff.&  Rose & Col. \\ \hline \hline
\multicolumn{5}{c}{SNR= 31 dB} \\ \hline
$\PSNR$& 46.8 & 48.2 & 47.8 & 46.7\\ \hline
$\SRDD\,\Nb=8$& 19.2 & 19.2 & 24.1 & 47.7 \\ \hline
Time (min) &1.6&1.6 &1.3  &0.9  \\ \hline
$\SRDD\,\Nb=16$&27.3&25.5& 38.7&110.6\\ \hline 
Time (min) &3.4 &3.8 &2.1 &1.1 \\ \hline
$\SRDD\,\Nb=24$&29.6&26.8& 44.2& 147.5\\ \hline 
Time (min) &7.6 &9.2 &4.5 &1.5\\ \hline
$\SRD\, \Nb=8$&49.1&59.7&74.6&137.2\\ \hline 
Time (min) & 18 &15 &10 &6 \\ \hline \hline
\multicolumn{5}{c}{SNR= 33 dB} \\ \hline
$\PSNR$& 48.8&50.2&49.8& 48.7\\ \hline
$\SRDD\,\Nb=8$&15.2&15.4&19.3 &41.5\\ \hline
Time (min) &2.3 &2.1 &1.7 &1.1 \\ \hline
$\SRDD\,\Nb=16$&20.4&19.5&29.1&86.4\\ \hline
Time (min) &5.4  &5.6 &2.9  &1.2 \\ \hline
$\SRDD\,Nb=24$& 21.9& 20.5& 32.7 & 106.3\\ \hline
Time (min) & 12 &  14  & 6.8 &1.9\\ \hline
$\SRD\,\Nb=8$& 33.5 &41.6& 53.2& 106.5\\ \hline
Time (min) &25&21 &16 &8 \\ \hline \hline
\end{tabular}
\caption{{\small{Values of SR for the approximation in the 
 pixel-intensity domain of the images listed in the 
 first row.  $\SRDD$ indicates the SR 
 for the plane by plane approximation in partition 
 of block side $\Nb=8,16$, and 24. $\SRD$ corresponds to a 
 partition in ${\mathrm{3D}}$ blocks of size $8\times 8 \times 8$. 
 The times for completing the approximations are given 
  immediately below the sparsity results in minutes.}}}
\label{TABLE3}
\end{center}
\end{table}

Every block in the partition is approximated up to the
same error. With all the approaches,
two global values of $\SNR$ (31 dB and 33 dB)
were considered.
These values of $\SNR$ correspond to
the values of $\PSNR$ shown in
 Tables ~\ref{TABLE3} and Table~\ref{TABLE4}.
In all of the cases the approximations are of excellent
visual quality.

 The SRs produced by the 3D approximation  
  are indicated  
 by $\SRD$ and those produced by the 2D plane 
 by plane approximation by $\SRDD$. 
 The times
 for completing the approximations are given
  in the row right after the corresponding sparsity result. 
\begin{table}[!h]
\begin{center}
\begin{tabular}{|l||c|c|c|c||}
\hline
Image & Ribei. & Graff.& Rose & Col.\\ \hline \hline
\multicolumn{5}{c}{SNR= 31 dB} \\ \hline
$\PSNR$&46.8& 48.2 & 47.8 & 46.7 \\ \hline
$\SRDD\,\Nb=8$ & 28.6& 26.8 & 38.6 & 56.5 \\ \hline
Time (min) & 1.4& 1.5 & 1.2  & 0.8 \\ \hline
$\SRDD\,\Nb=16$& 36.5& 34.1 & 63.4 & 144.8  \\ \hline
Time (min) & 2.7& 3.5 & 2.3 &  0.9\\ \hline
$\SRDD\,\Nb=24$& 37.2 & 35.7& 71.1  & 193  \\ \hline
Time (min) & 9.2  & 12 & 4.8  & 1.8 \\ \hline
$\SRD\,\Nb=8$ &86.5&108.0& 182.2& 371.7\\ \hline 
Time (min) & 13 & 10  & 6 & 3  \\ \hline \hline
\multicolumn{5}{c} {SNR= 33 dB} \\ \hline
$\PSNR$& 48.8& 50.2 & 49.9 & 48.7 \\ \hline
$\SRDD\,\Nb=8$& 22.6 & 21.8 & 33.0 & 56.1 \\ \hline
Time (min) &1.7 &1.8& 1.5 & 1.0 \\ \hline
$\SRDD\,\Nb=16$& 26.6& 25.8 & 48.2 & 118.3 \\ \hline
Time (min) &3.5&5.0& 2.3&1.1 \\ \hline
$\SRDD\,\Nb=24$&21.9&26.8&52.0&144.0\\ \hline
Time (min) &12&15&8.5&1.9 \\ \hline
$\SRD\,\Nb=8$& 55.1&70.5&129.5& 313.3\\ \hline
Time (min) &23&18&10&1.8\\ \hline \hline
\end{tabular}
\caption{{\small{Same description as in Table ~\ref{TABLE3}, 
but the approximations are realized by applying first 
a wavelet transform to each of the 32 channels.}}}
\label{TABLE4}
\end{center}
\end{table}

{\em{Remark 3:}}
In both Table~\ref{TABLE3} and 
Table~\ref{TABLE4} the values of $\SRD$ 
are significantly larger than the values of $\SRDD$, 
 except for the Col. image
and $24 \times 24$ blocks. For this image we were 
able to increase the 3D block size up to $16 \times 16 \times 16$ and the results for $\SNR=31$dB 
are $\SRD=357$ in the pd and $\SRD= 892$ in the wd
(35 min and 10 min respectively). 
For $\SNR=33$ dB $\SRD=247$ in the pd and $\SRD=590$ 
in the wd (55 min and 20 min respectively).
 
On comparing the two tables a drastic improvement 
in the values of $\SRD$ is observed when the approximation 
is realized in the wavelet domain. This feature is 
a  consequence of the fact that the planes of 
the natural images  are very sparse in the wavelet domain.
  In order to highlight differences  we 
  produce next the $\SRD$ corresponding 
 to the two remote sensing images in Fig.~\ref{RS}. 
The graph on the left  represents the   
 Urban remote sensing hyper-spectral 
 image taken from \cite{HSRS}. The graph on the right 
 is a portion of the University of Pavia 
image also taken from \cite{HSRS}. 
\begin{figure}[ht]
\centering
%\vspace{-0.9cm}
%\includegraphics[scale=0.74]{urban.eps}\hspace{0.4cm}
\includegraphics[scale=0.666]{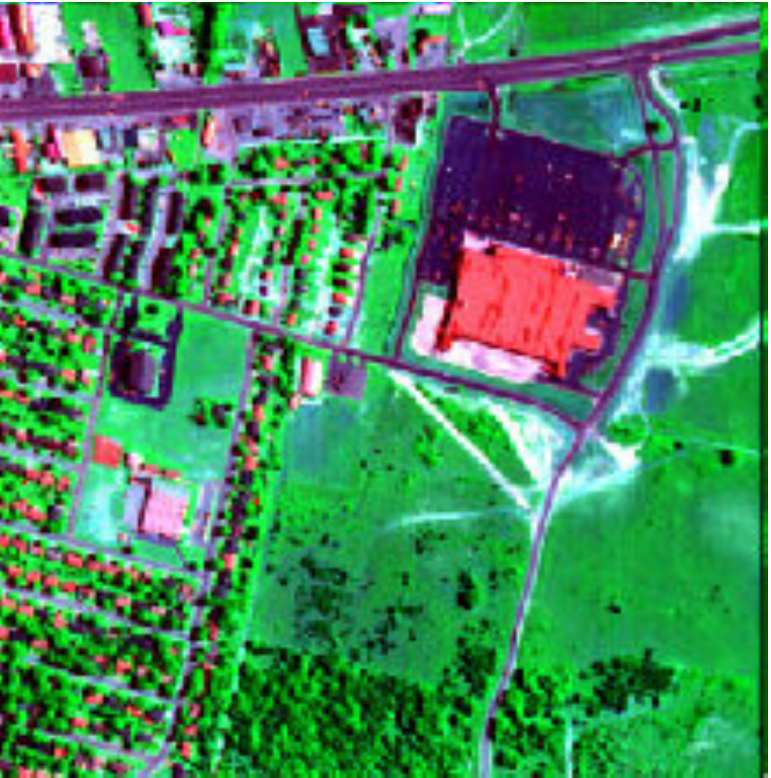}\hspace{0.4cm}
\includegraphics[scale=0.774]{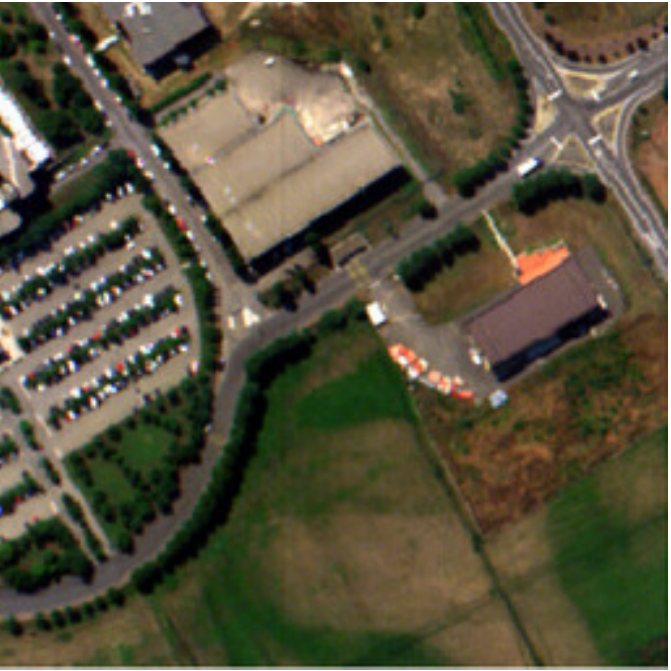}
\caption{{\small{Illustration of two remote sensing 
hyper-spectral images taken from \cite{HSRS}.
The graph on the left is the Urban 
image (size $320 \times 320 \times 128$ pixels). 
 The graph on the right is a portion of the 
University of Pavia image 
($256 \times 256 \times 96$ pixels).
}}}
\label{RS}
\end{figure}

 Fig.~\ref{SRRS} plots  the SR 
 vs four values of $\SNR$, 
corresponding to the 3D approximations of 
the Urban and  University of Pavia images 
in both the pd and wd.
\begin{figure}[ht!]
\centering
%\vspace{-0.9cm}
\includegraphics[width=9cm]{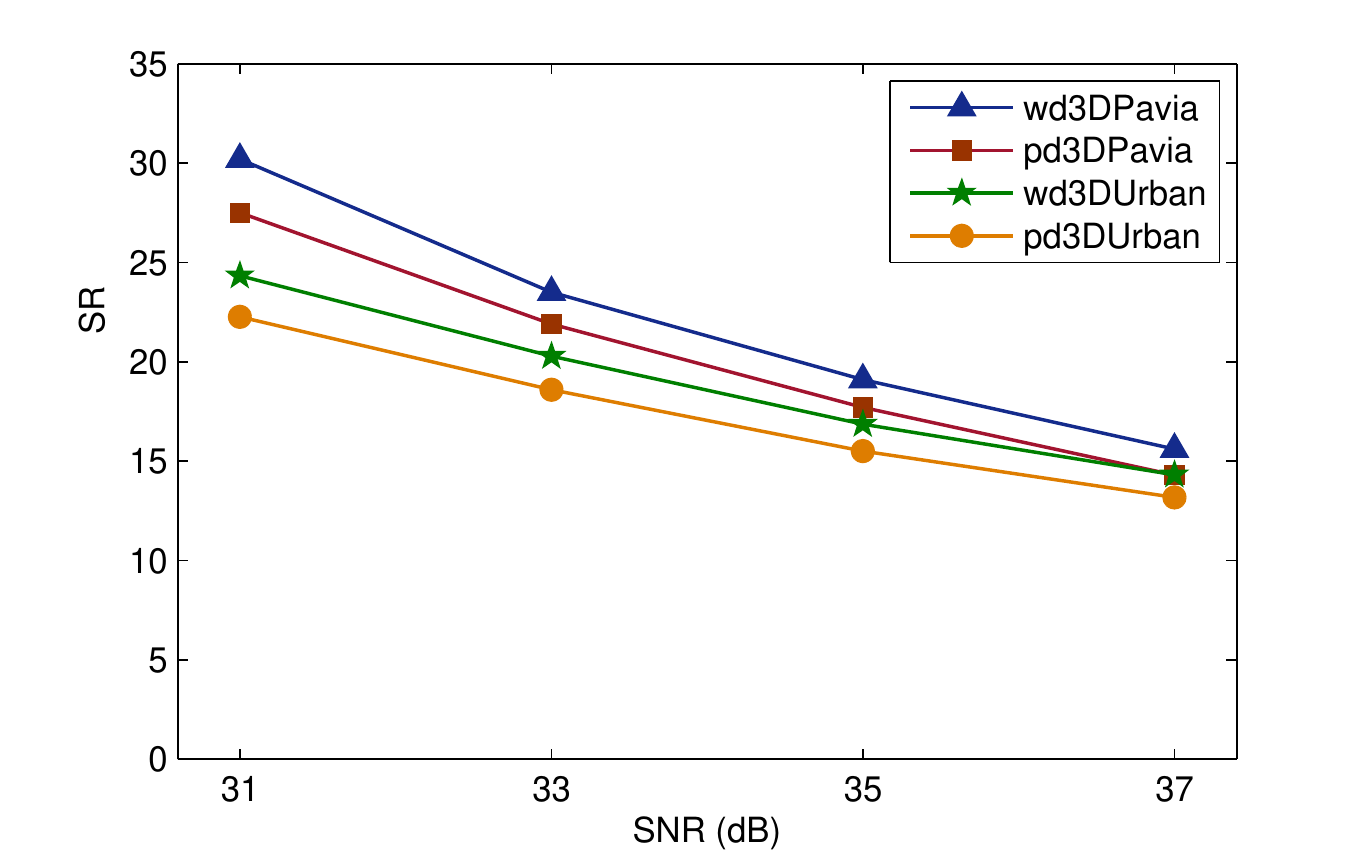}
\caption{{\small{SR vs SNR values for the 
3D  approximation in 
both the pd and wd for the Urban and University of Pavia 
 remote sensing images.}}}
\label{SRRS}
\end{figure}

Notice that the results in the pd are much closer to the results in the wd
than they are in the case of the natural images
 in Fig.~\ref{hsi}.  This is because, as 
illustrated in Fig.~\ref{WT}, the planes of   
the remote sensing  images are not as 
 sparse in the wd as the planes of the natural images are.
\begin{figure}[ht]
\centering
%\vspace{-0.9cm}
\includegraphics[width=8cm]{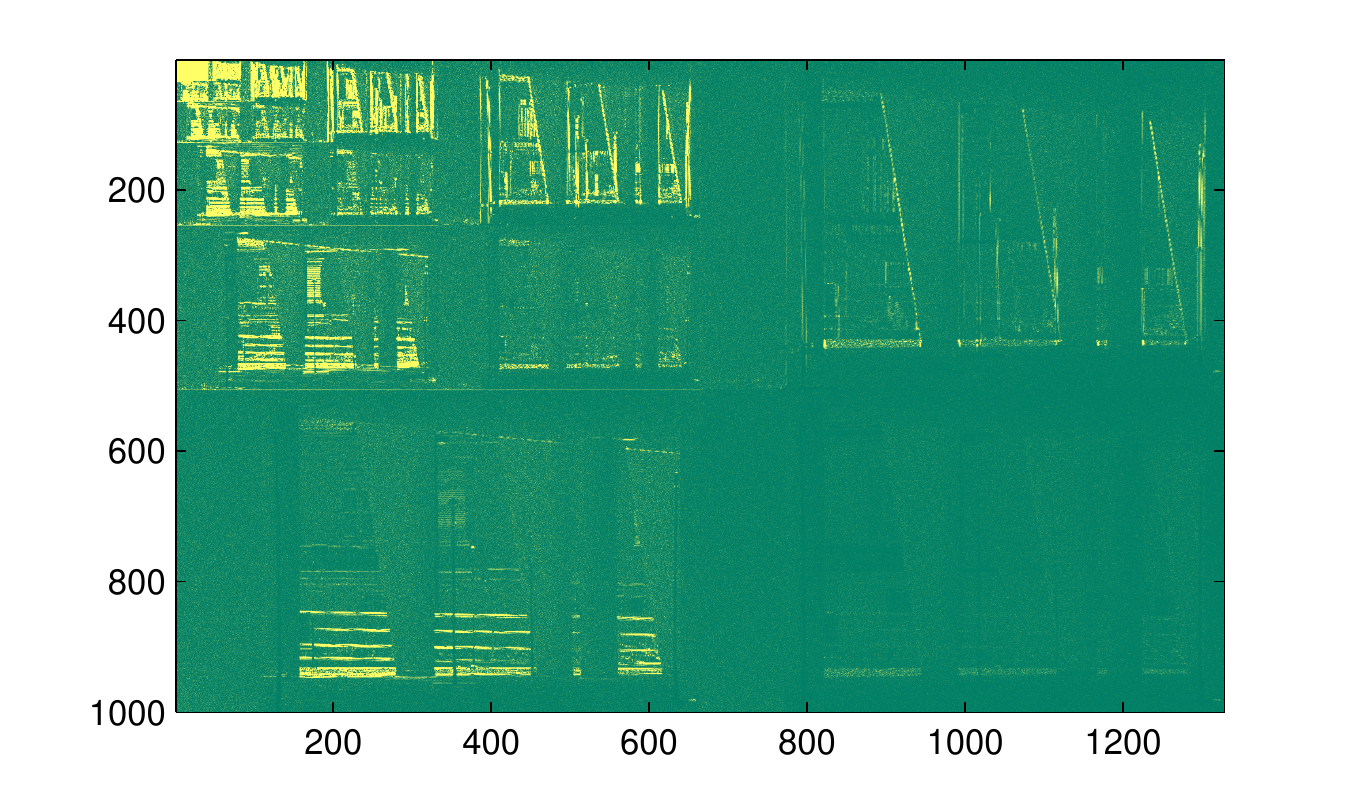}
\includegraphics[width=8cm]{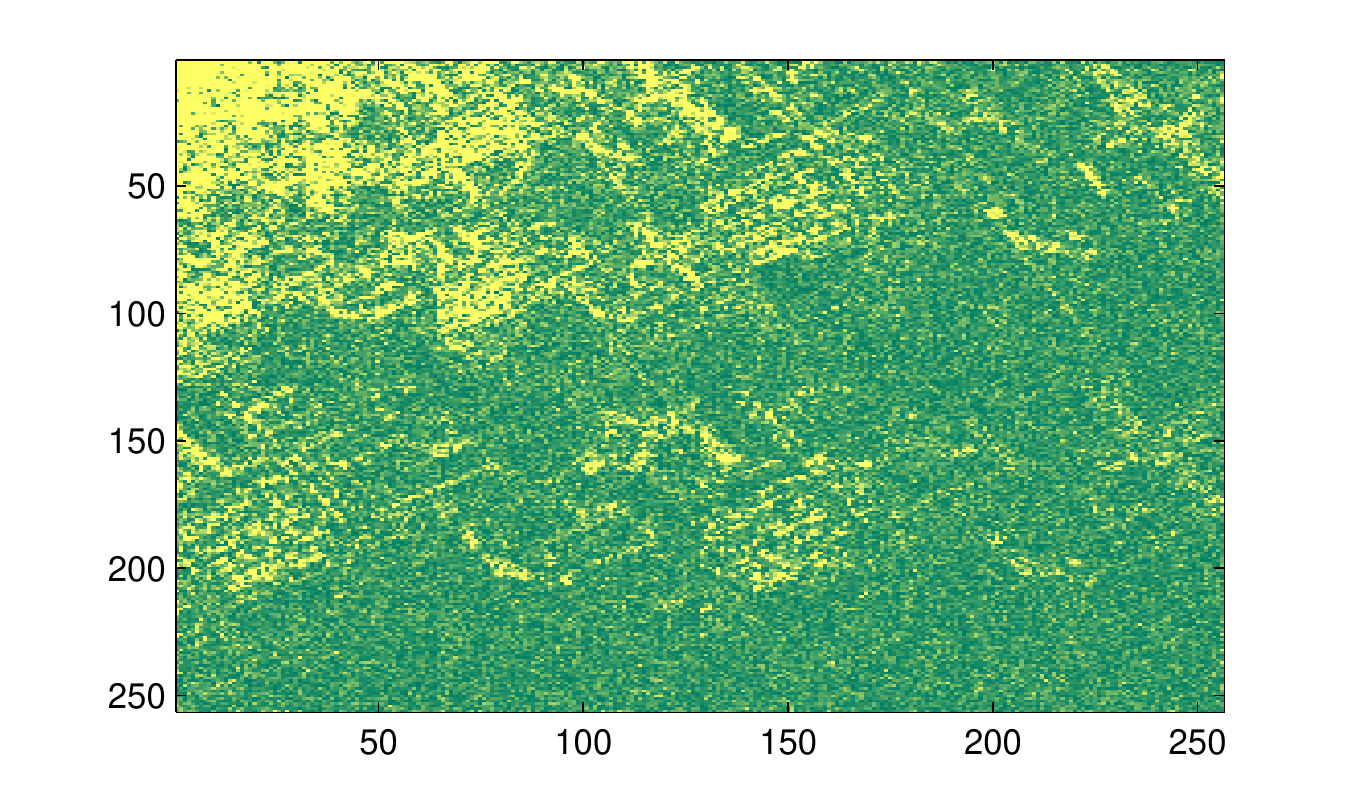}
\caption{\small{Absolute value of the wavelet transform 
of a plane in the Col. image  (left graph) and   
in the University of Pavia image (right graph).}}
\label{WT}
\end{figure}
\newpage

\section{Conclusions}
High quality approximation of 3D images has been considered   within the context of data reduction. A remarkable
improvement in sparsity achieved by the simultaneous
approximation of multiple channels has been illustrated 
  through numerical experiments of different natures.
Firstly it was demonstrated that   
a standard data set of RGB images can be approximated at
high quality using far fewer  elementary components 
 if each image is treated as a very thin 3D array
instead of as 3 independent 2D arrays. 
  Secondly the
full power of the approach was demonstrated  through the
approximation of hyper-spectral images. For the 
 hyper-spectral natural images the sparsity is  remarkably 
 higher if the  approximation is realized in the 
wavelet domain. 
 For the remote sensing images 
 the domain of approximation has less influence because,
 as opposed to natural images, these images are not 
 as sparse in the wavelet domain as natural images are.

Taking into account the major reduction of dimensionality 
 demonstrated by the numerical examples in this work, 
we feel  confident that the proposed approach will be 
of assistance to the broad
range of image processing applications which rely
on a transformation for data reduction as a first step
of further processing.
\subsection*{Acknowledgments}  
 We are indebted to P. Getreuer,  
 for making available the {\tt{waveletcdf97}}  
  MATLAB function that we have used for the 
transformation of each single channel image to the wavelet 
domain.

\end{document}